\documentclass[fleqn,usenatbib]{mnras}

\usepackage[T1]{fontenc}
\usepackage{xcolor}

\DeclareRobustCommand{\VAN}[3]{#2}
\let\VANthebibliography\thebibliography
\def\thebibliography{\DeclareRobustCommand{\VAN}[3]{##3}\VANthebibliography}

\usepackage{graphicx}	
\usepackage{amsmath}	
\usepackage{tablefootnote}
\usepackage{threeparttable}
\usepackage{amssymb}

\defcitealias{Nardini}{Paper 1}

\newcommand{\kms}{\,km\,s$^{-1}$}	

\title[Pulsating B-type stars in young clusters]{Characterisation of B-type stars in four young Galactic open clusters. \\ II. Photometric variability with TESS and identification of pulsating stars}

\author[F. Nardini et al.]{F. Nardini$^{1}$,
D. M. Bowman$^{1,2}$,
J. Bodensteiner$^{3}$, 
P. Van Daele$^{1}$,
L. J. A. Scott$^{1}$, 
K. Thomson-Paressant$^{1}$,
\and
J. Henneco$^{1}$,
H. Sana$^{2,4}$
\\
$^{1}$ School of Mathematics, Statistics and Physics, Newcastle University, Newcastle upon Tyne, NE1 7RU, United Kingdom\\
$^{2}$ Institute of Astronomy, KU Leuven, Celestijnenlaan 200D, 3001 Leuven, Belgium\\
$^{3}$ Anton Pannekoek Institute for Astronomy, University of Amsterdam, 1090 GE Amsterdam, The Netherlands \\
$^{4}$ Leuven Gravity Institute, KU Leuven, Celestijnenlaan 200D, 3001 Leuven, Belgium\\
}

\date{Accepted XXX. Received YYY; in original form ZZZ}

\pubyear{\the\year{}}

\begin{document}
\label{firstpage}
\pagerange{\pageref{firstpage}--\pageref{lastpage}}
\maketitle

\begin{abstract}
Pulsations and binarity are important properties of massive stars and influence their structure and evolution. Young open clusters are ideal laboratories to study massive stars, since they are mostly coeval and chemically homogeneous. We analyse 82 B-type stars in the Galactic open clusters $h$ and $\chi$ Persei, NGC 457, NGC 581, and NGC 1960, which have ages between 15 and 30 Myr. We extract light curves using both simple aperture photometry and a point-spread-function-based method that minimises contamination from crowded fields in full-frame images from the Transiting Exoplanet Survey Satellite (TESS) mission. We successfully extract light curves for 63 stars. The majority have photometric variability caused by pulsations, showing how pulsations are a common property of young B-type stars in clusters. We classify stars into different pulsator types based on their spectral type and frequency spectrum. The most common pulsator class are beta Cephei stars, with 29 stars in total, and we discover pulsation modes in almost all stars with a spectral type of B2 or earlier. All Be stars show pulsations, which supports the consensus of the ubiquity of pulsations in these stars. Finally, we confirm that the majority of early B-type stars are in binary systems, except for two pulsating Be stars and a star with rotational modulation. 
This indicates that pulsations do not significantly bias the spectroscopic detection of short-period binaries, except for Be stars where the spectroscopic radial velocity variability threshold should be higher than the typical value of 20 \kms.

\end{abstract}




\section{Introduction}

In recent years, our understanding of the physical processes that govern stellar structure and evolution has largely improved \citep[see][for a monograph]{Kippenhahn}. However, we still lack a sufficient understanding to predict the end products of single-star evolution. This is because of the variety of physical processes in their lives, from strong stellar winds, high rotation rates that influence the internal mixing, and potentially magnetic fields \citep{langer2012}. For example, large theoretical uncertainties for mixing exist in stellar evolution models, both due to rotation and other instabilities related to convective boundaries, which play a crucial role in the mass of the helium core, the main-sequence lifetime, and subsequent evolution \citep[][]{Talon_1997, Johnston_2021}. Moreover, most massive stars are born in binary or multiple systems and eventually interact with at least one companion \citep{sana2012, sana2013, deMink_2014, Offner_2023, Frost_2024, Bordier_2024}. These interactions, such as mass transfer and tides, introduce additional complications and can even lead to mergers \citep[see][for recent reviews]{Marchant_bodensteiner, Schneider_2025}.

Asteroseismology is a powerful tool that allows us to constrain the structure of stars across the Hertzsprung--Russell (HR) diagram using stellar pulsations \citep{Aerts_2010}. For example, the asteroseismic analysis of massive stars has allowed us to estimate their interior rotation and mixing profiles \citep{Aerts_2003a, Dupret2004, Burssens_2023, Vanlaer_2025, Fritzewski_2025}, and even constrain their interior magnetic fields \citep[e.g.][]{Lecoanet_2022, Vandersnickt_2025}. 
Modern space telescopes, such as MOST \citep{Walker_MOST}, CoRoT \citep[][]{Auvergne_corot}, BRITE \citep[][]{Weiss_brite}, Kepler/K2 \citep[][]{Borucki_kepler, Howell_k2}, and the ongoing TESS mission \citep[][]{Ricker2015} have shown that almost all massive stars in the Galaxy pulsate \citep[e.g.][]{Balona2011, Burssens_2020, Thomson-Paressant_2026}. Moreover, a large fraction of massive stars in the low-metallicity Large and Small Magellanic Clouds exhibit photometric variability that probes their mass and age \citep{Bowman_2019_slf, Bowman_2024, Linhao_2024}. 
Massive stars are intrinsically rare compared to low-mass stars, and these space missions have provided the first high-precision, long-duration time-series data for a large sample of massive stars suitable for ensemble asteroseismology \citep[see][for a review]{Bowman_2020}. For massive stars in binary systems, which can be subject to mass transfer, mergers, and tidal forces, asteroseismology is more complicated but provides valuable constraints on binary evolution theory \citep[see][for a review]{Southworth_Bowman_2026}.

Pulsations are typically characterised by their dominant restoring force. Gravity modes have buoyancy as the restoring force, while pressure modes have the pressure gradient as their main restoring force.
In massive stars, pulsations can be excited by different mechanisms, including a heat-engine mechanism operating in local opacity bumps ($\kappa$ mechanism) near the stellar surface \citep{Dziembowski_betac, Dziembowski_spb, Miglio_2007}, stochastically excited gravity waves \citep[see][for a review]{Bowman_2023} and tides in close binary systems \citep[][]{Southworth_Bowman_2026}.
It is sometimes difficult to identify the pulsation excitation mechanism from observations alone, since this requires non-adiabatic pulsation models (e.g. \citealt{Townsend_2013}). However, since the $\kappa$ mechanism excites coherent pulsation modes with long lifetimes and turbulent convection typically excites damped modes with short lifetimes, it is usually possible to distinguish between the two cases.

Among stars of spectral types O and B, there are three main pulsator types: slowly pulsating B-type stars (SPB), $\beta$~Cephei stars, and stochastic low-frequency (SLF) variables. These are separated because of distinctions in their excitation mechanism and/or instability regions in the HR~diagram. SPB stars have spectral types between B3 and B9 and typically have high-radial order gravity modes excited by the $\kappa$ mechanism \citep{Dziembowski_spb, Papics_2017, Szewczuk_2018, Szewczuk2021, Pedersen_2021, Sharma_2022}.
At higher masses, the $\beta$~Cep stars have birth masses approximately between 8 and 30 M$_{\odot}$ and typically show low-radial order pressure and gravity modes \citep{Dziembowski_betac, Stankov_2005, Burssens_2023, Fritzewski_2025}. The pulsation periods of SPB and $\beta$~Cep stars are of order several hours to a few days. 
It is also possible to have so-called ``hybrid'' pulsators \citep{Handler2009}, which are stars that show characteristics of different pulsator classifications. Historically, this has been used to refer to stars with characteristics of both SPB and $\beta$~Cep stars and located in the overlapping instability regions of the HR~diagram.

Another relevant group of pulsators among early-type stars are Be stars. 
Space-based photometry has demonstrated that non-radial pulsations are almost ubiquitous among Be stars \citep[][]{Rivinius_2003, semaan, Labadie_Bartz_2022}. Both the light curves and the frequency spectra of pulsating Be stars are usually characterised by similar properties \citep[e.g. complex beating patterns and frequency groups;][]{Bradley_2015, Kurtz_2015}. 
On longer time scales of months to years, Be stars experience epochs of brightening or fading, which are likely attributed to the dissipation or growth of an equatorial decretion disc created by mass ejections \citep{Rivinius, Rivinius_2016, semaan, Labadie_Bartz_2025_disk, Nova_2025}. 
Spectroscopically, classical Be stars are characterised by rapid rotation as well as emission lines in their spectra caused by the presence of the equatorial decretion disc \citep[][]{Rivinius, Rivinius_2026}. The formation of a Be star, specifically if it was born as a single fast rotator or if it is a binary interaction product, is still debated in the literature \citep[][]{Hastings2020, Bodensteiner_2020_lackbe}.

The majority of massive stars are born in clusters or associations, and a major advantage of investigating binarity and pulsations is that knowing the cluster age and assuming coeval evolution for the member stars provides a significant advantage, particularly when one of the goals is to identify candidates for pre- and post-interaction binary systems. 
This reduces the parameter space in subsequent modelling. Comparing multiple clusters spanning a range of ages allows an investigation of how the binary and pulsation properties of stars change in time. Recently, \citet{Nardini} (hereafter \citetalias{Nardini})  
performed a spectroscopic characterisation of a sample of 74 early B-type stars in four young open clusters: $h$ and $\chi$ Persei (NGC 869 and NGC 884, respectively), NGC~457, NGC~581, and NGC~1960. The clusters were selected based on their ages to fill a gap in the literature from (very) young to intermediate age: 15~Myr for $h$ and $\chi$~Persei; 24~Myr for NGC~457; 29~Myr for NGC~581; and 30~Myr for NGC~1960 \citep{Dias2021}. High-resolution multi-epoch HERMES \citep{raskin_2011} spectroscopy was used to identify binary systems based on radial velocity (RV) variability, revealing no statistically significant differences in the binary fraction of the four clusters, with an overall bias-corrected binary fraction of $79^{+19}_{-16}$~per~cent (see \citetalias{Nardini}). 

In this second paper, we characterise the photometric variability of the same clusters using time-series space photometry from the NASA TESS mission \citep{Ricker2015}, with a particular focus on identifying pulsating stars.
In Sect.~\ref{sec:Methods}, we describe the sample, discuss the methods used to extract the light curves and perform the frequency analysis, and examine the differences in the atmospheric parameters determined using two different methods. Section~\ref{sec:variability in clusters} describes the variability in the four clusters with a focus on pulsating Be stars, $\beta$ Cep, SPBs and hybrids. Section~\ref{sec:binary systems} revisits the binary systems and discusses the photometric binaries identified in the sample. Lastly, we give a summary and conclude in Sect.~\ref{sec:conclusion}.

\section{Methods}\label{sec:Methods}
\subsection{Sample selection}
\begin{figure}
    \centering
    \includegraphics[width=0.99\linewidth]{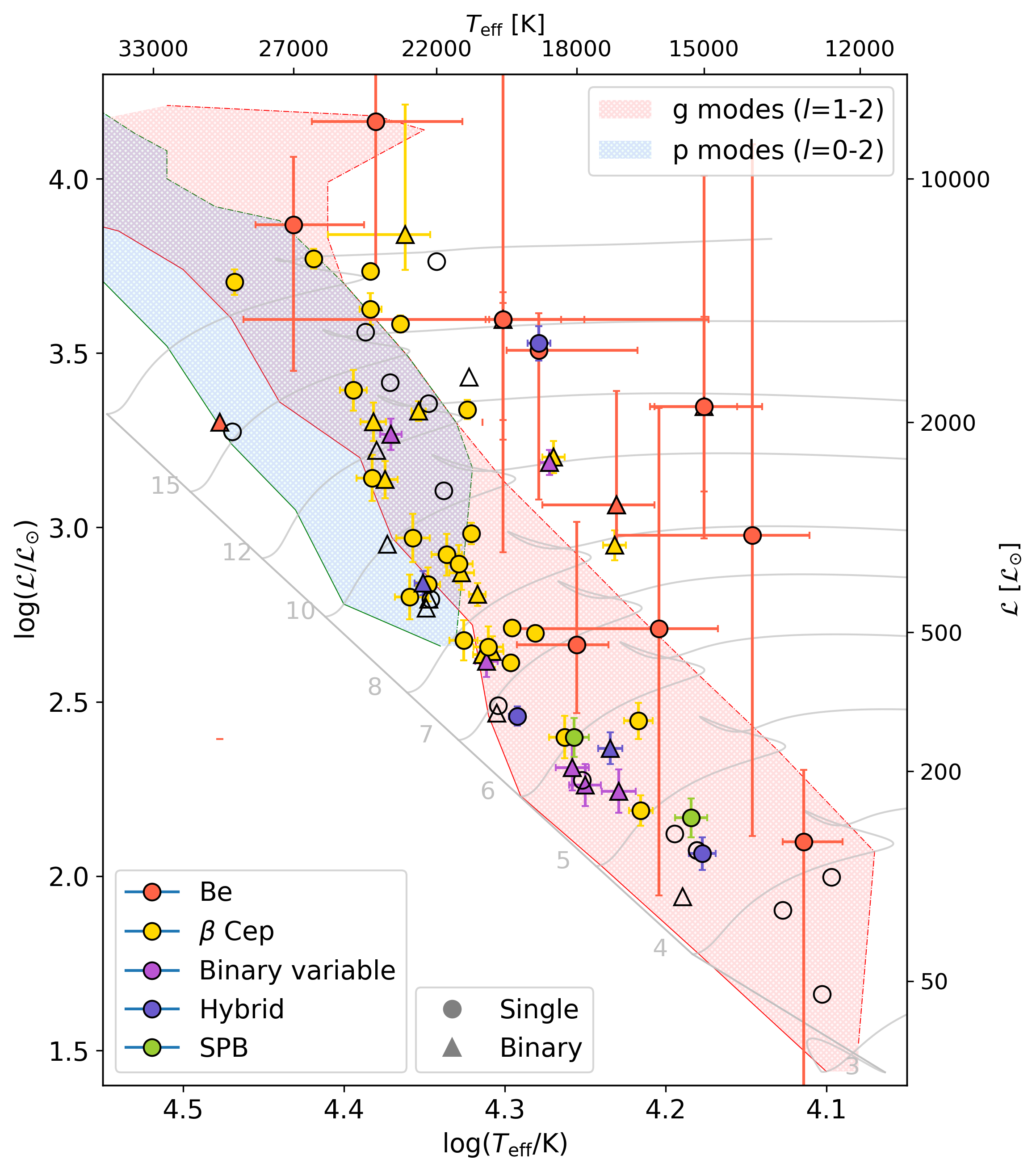}
    \caption{Asteroseismic HR~diagram of the combined sample of 82 B-type stars across the four young open clusters studied in this work. We represent different classes of pulsators with different colours, and the markers indicate apparent single stars (circles) and binaries (triangles). Evolutionary tracks in units of M$_{\odot}$ and theoretical pulsation instability strips are from \citet{Burssens_2020}, for which the blue shaded region is for pressure modes, and red is for gravity modes. The stellar parameters for non-emission line stars were determined using Gaia's ESP-HS module, whereas these were determined using a grid of {\sc tlusty} \citep[][]{tlusty_1995, tlusty_2007} models for the Be stars.}
    \label{fig:HRD}
\end{figure}

In this paper, we analyse a sample of 82 stars in four young open clusters: 26 are members of $h$ and $\chi$ Persei, 19 are from NGC~457, 14 are from NGC~581, and 23 are from NGC~1960. This sample is slightly larger than the 74 B-type stars analysed spectroscopically in \citetalias{Nardini}. Here, we included the additional three members of the cluster $h$ and $\chi$ Persei (NGC~869~0622, BD+56~515 and NGC~884~2114) that lacked multi-epoch HERMES spectroscopy for the binary analysis in \citetalias{Nardini}. 
Additionally, five late-B stars (NGC~1960~3, NGC~1960~12, NGC~1960~14, NGC~1960~86 and NGC~1960~92), which span spectral types from B7 to B9, were not included in \citetalias{Nardini} but included in this work to ensure better coverage of the full mass range of B-type stars in all four clusters. Figure~\ref{fig:HRD} shows an asteroseismic HR diagram for all 82 stars in our combined four-cluster sample. The properties of the stars are reported in Table~\ref{tab:stars_list} (see also Sect.~\ref{sec:tlustyvsgaia}).

We note that the precise location of the main-sequence in Fig.~\ref{fig:HRD} depends on the input physics and numerics used in theoretical evolutionary tracks, which affects any interpretation of whether a star can be considered to be on the main sequence. In our case, the evolutionary tracks shown in Fig.~\ref{fig:HRD} were computed by \citet{Burssens_2020} using the Modules for Experiments in Stellar Astrophysics \citep[MESA;][v.12115]{paxton1, paxton2, paxton3, paxton4}. These tracks assume non-rotating stars with Galactic metallicity $Z_{\rm ini} = 0.014$, an initial hydrogen abundance of $X_{\rm ini} = 0.71$, a diffusive exponential profile for convective boundary mixing (CBM) with $f_{\rm CBM} = 0.02$, and an envelope mixing coefficient of $\log D_{\rm mix} = 1.0$.
Different implementations for rotation and mixing (e.g. more efficient mixing due to faster rotation) would shift the terminal-age main-sequence (TAMS) towards cooler temperatures and increase the main-sequence lifetime (see \citealt{Bowman_2020}).

\subsection{TESS light curve extraction}\label{sec:lc extraction}
The NASA TESS space mission is an all-sky survey which had a primary-mission objective to detect exoplanets transiting bright stars \citep[][]{Ricker2015}. The field of view is $24^{\circ} \times 96^{\circ}$ and TESS observes the entire sky in a series of observational sectors that last for at least 27~d. Originally launched in 2018, TESS is currently in its third extended mission and has returned to re-observe stars in multiple sectors, meaning that the total time span of its assembled light curves spans several years in most cases. This survey approach facilitates the discovery and follow-up asteroseismic analysis of many massive stars (see review by \citealt{Bowman_2020}).

Although working with open clusters provides the advantage of having a stellar sample of the same age and chemical composition, crowded fields are a major challenge when it comes to light curve extraction with the TESS mission (e.g. \citealt{Bowman_2024, VanDaele_2026}). Due to TESS's large pixel size (21 arcsec), blending and contamination are often inevitable for dense environments such as clusters. However, these limitations depend on the location of a star within the cluster and the proximity and brightness of other stars, thus each target has a different level of contamination.

\subsubsection{Simple aperture photometry}
\begin{figure*}
    \centering
    \includegraphics[width=0.99\linewidth]{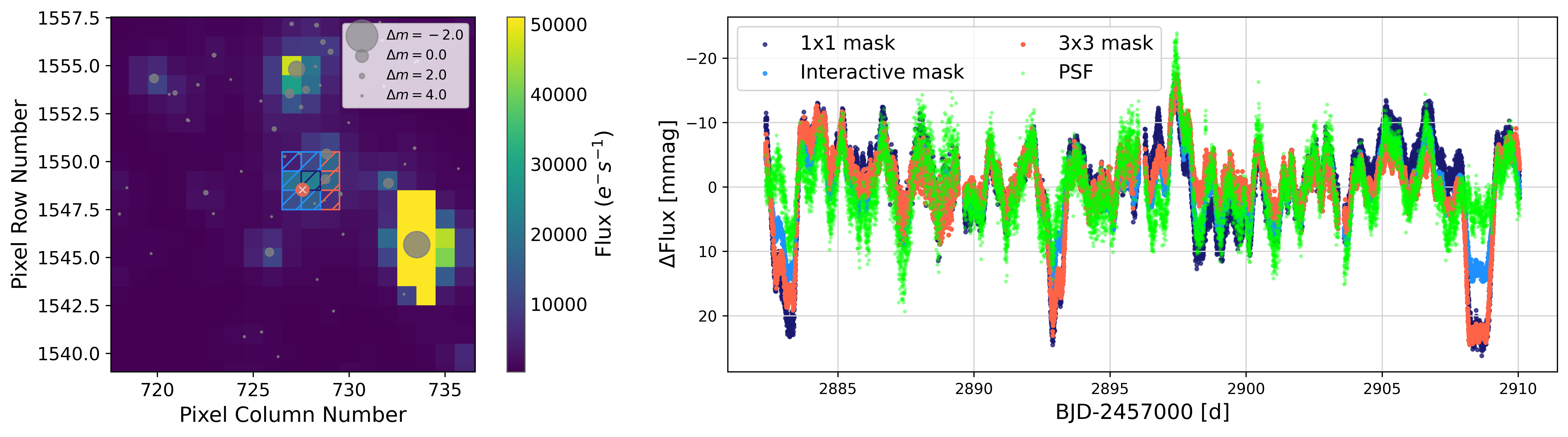}
    \caption{Example of the light curve extraction methodology for sector~58 of NGC~884~2296. The left panel shows the $20\times20$ pixel cut-out of the TESS FFI with the target star in the centre. The different coloured hatched areas are the different aperture masks: $3\times3$ (red), interactive (light blue), and $1\times1$ (dark blue). The position of the target and neighbouring stars was determined using the {\tt Python} package \textsc{tpfplotter} \citep[\url{https://github.com/jlillo/tpfplotter};][]{tpfplotter}, which overplots the Gaia DR3 catalogue over the TESS FFI cut-out, and the size of the points is proportional to the TESS magnitude difference. The right panel shows the different light curves extracted with the different aperture masks in the corresponding colours. Additionally, the PSF light curve extracted using the {\sc lemons} software \citep{VanDaele_2026} is shown in green, which demonstrates that the eclipses do not originate from the target under investigation.}
    \label{fig:example_diagnostic}
\end{figure*}

TESS full-frame images (FFIs) are available with cadences of 30~min, 10~min, and 200~s for the nominal, first-extended and second-extended cycles, respectively, and are publicly available on MAST\footnote{\url{https://archive.stsci.edu/missions-and-data/tess}}.
At the time of our analysis, TESS FFIs were available up to sector 85 for the clusters $h$ and $\chi$ Persei and NGC~581, to sector 58 for NGC~457, and to sector 73 for NGC~1960.
We used the \textsc{lightkurve}\footnote{\url{https://github.com/lightkurve/lightkurve}} package \citep{Lightkurve_coll} together with \textsc{tesscut} \citep{Brasseur_2019} to download $20\times20$ pixel cutouts of the FFIs for each target. To extract light curves, we used simple aperture photometry (SAP) with three different aperture masks: a $1\times1$ pixel mask, a $3\times3$ mask, and an optimised mask selected interactively for each target by visual inspection to maximise the signal-to-noise ratio (S/N) of variability and minimise contamination (see discussion by \citealt{ScottBowman2026}). The $1\times1$ pixel mask is a useful diagnostic tool to check if any detected variability originates from the target or from a neighbouring contaminating source. Meanwhile, whether the $3\times3$ or the interactive masks were better and thus used for subsequent analysis depends on the level of blending and contamination of each target. An example of the three different masks for NGC~884~2296 and the resultant light curves is shown in Fig.~\ref{fig:example_diagnostic}. 

To reduce the raw TESS light curves, a background mask is constructed to include all pixels with flux levels below 0.5~per~cent of the target star’s flux, as these pixels contain mostly background light. These background-corrected light curves are then detrended using principal component analysis \citep[PCA;][]{Hotelling_pca}.
Detrending using PCA identifies and removes instrumental trends and systematics in the background-corrected light curves. The optimal number of principal components was determined by trialling different options and inspecting their behaviour over time. We selected only those components that did not display any periodic signals, ensuring that the detrending process would not introduce or remove any astrophysical variability in the output light curves. This approach has proven reasonable for massive stars (see e.g. \citealt{Bowman_cubespec, ScottBowman2026}). Finally, additional detrending was applied to all stars using a long-period spline fit with the {\tt Python} \textsc{splrep} and \textsc{splev} functions from \textsc{scipy}\footnote{\url{https://docs.scipy.org/doc/scipy/reference/interpolate.html}} to remove any sector-long trends missed by the PCA detrending.

\subsubsection{Contamination in crowded fields}
Light curve extraction is straightforward for targets that are isolated or only have nearby stars that are several magnitudes fainter. In these cases, using a $3\times3$ pixel mask was generally the best approach. However, for targets with a contaminating star of similar brightness within a few pixels, the interactive mask was deemed the best option since we can then manually select the pixels that maximise the target's flux and minimise the contaminating flux. 

Table~\ref{tab:stars_list} includes a summary of the TESS light curve extraction results and the contamination ratio ($C_{\rm ratio}$) for all stars. This quantifies the amount of contamination based on the \texttt{Python} code \textsc{tic inspect}\footnote{\url{https://github.com/mpaegert/tic_inspect}}, and is defined by \citet{Stassun_2018} as $C_{\rm ratio} = F_{\rm n}/F_{\rm s}$, where the amount of flux from the neighbouring sources ($F_{\rm n}$) is divided by the flux of the target star ($F_{\rm s}$). 

For 18 of our targets, there is a star of equal brightness or brighter one pixel away from the central pixel. Therefore, even the $1\times1$ pixel mask option yields a large amount of contamination. This means that for such cases, we were only able to tentatively assign them to a pulsator class, but we did not perform any further analysis. These objects are reported in Table~\ref{tab:stars_list} with daggers next to their variability class.
Lastly, it was impossible to robustly extract a light curve at all for 19 targets, since another brighter star is located in the same pixel. The median of the contamination ratio for these targets is 4, while for the 18 tentative targets it is 2.3.

\subsubsection{PSF fitting}

For stars that were not completely isolated, but the contamination was not too high, we used the {\sc Lemons} software developed by \citet{VanDaele_2026}.
{\sc Lemons}\footnote{\url{https://github.com/pieterjanv314/Lemons}} performs light curve extraction using point spread function (PSF) fitting of TESS FFIs. In this approach, the flux of the target is extracted as a weighted sum of the pixel fluxes, where each pixel's weight is assigned according to the best-fit PSF assuming a 2D Gaussian. The \textsc{Lemons} software uses Gaia positions to locate stars and is further optimised for early-type stars since it allows for variable centroid and broadness as fitting parameters for the target and nearby contaminating stars. This is important since the broadness of the PSF is time-dependent for a variable star \citep{VanDaele_2026}.
Furthermore, by allowing the centroid to vary from frame to frame, one compensates more robustly for the drift of the TESS spacecraft, which causes significant smearing of flux across the CCD during a TESS sector (see discussion by \citealt{VanDaele_2026}). Without mitigating this effect, PSF light curves include artificial long-term trends, especially for fainter stars ($T_{\rm mag} \gtrsim 12$). 

Figure~\ref{fig:example_diagnostic} includes an example in which the PSF light curve successfully demonstrates how contamination from a nearby eclipsing binary occurs for all of the $1\times1$, $3\times3$, and interactive pixel masks in the SAP methodology. In this particular example, all light curves obtained using the SAP masks show eclipses, which would lead to an incorrect classification for the target NGC~884~2296. Using the PSF fitting, we mitigate the contaminating flux from the nearby eclipsing binary.

\subsection{Photometric variability}\label{sect:variability classification}

Once the optimum light curve for each target had been determined from the SAP or PSF methods, we computed the discrete Fourier Transform (dFT) using the \textsc{Period04}  software \citep[][]{period04}. We calculated a dFT for each individual sector for each target to determine if the dominant frequencies were consistent across all sectors. Moreover, the large gaps (> 2 yr) between TESS cycles make frequency analysis of a combined light curve extremely challenging because of the complex spectral window (see \citealt{ScottBowman2026}). 

\subsubsection{Variability classification}

The different definitions of variability classes applied to our extracted TESS light curves include: 
\begin{enumerate}
    \item $\beta$~Cephei stars are dwarfs and giants with a spectral type of B2 or earlier and pulsation frequencies above about 4~d$^{-1}$ \citep[see ][]{Stankov_2005}; 
    \item SPB stars are dwarfs with spectral type B3 or later and show pulsations below 4~d$^{-1}$ \citep[][]{Decat_2007};
    \item Hybrids are dwarfs and giants with a spectral type later than B3 and show pulsations both above and below 4~d$^{-1}$;
    \item Pulsating Be stars are dwarfs showing emission in their spectra and pulsations that typically form frequency groups in the amplitude spectrum \citep[see][]{Kurtz_2015};
    \item Photometric binaries are targets that show either eclipses or ellipsoidal variability in their light curves due to binarity \citep[see][]{Southworth_Bowman_2026};
    \item Rotational modulation stars show periodic modulation in their light curves commensurate with their rotation period, which is likely caused by chemical abundance spots \citep[e.g. ][]{Bowman2018, Balona_2019};
    \item $\delta$ Scuti stars are those with with spectral type later than B7 and that show significant pulsation frequencies above 40~d$^{-1}$ \citep[see ][]{BowmanKurtz2018};
    \item constant stars show no signs of variability.
\end{enumerate} 
Note that these classification categories are not mutually exclusive. For example, some photometric binaries also exhibit pulsational variability. However, in this work, we classify stars based on their dominant photometric variability. The results of our photometric variability classification are shown in Table~\ref{tab:stars_list}.

Variability classification was performed for the subset of 63 stars that are not heavily contaminated ($C_{\rm ratio}\le 1$), with the frequency range of observed variability and the spectral type being the main criteria to distinguish different categories. For the remaining 19 stars, the contamination was deemed too high to reliably determine their variable classification. We discuss each class of pulsators in Section~\ref{sec:variability in clusters}, and discuss the photometric binaries in Section~\ref{sec:binary systems}. 

\subsubsection{Inclination-radius-velocity relation}

For stars that show rotational modulation, we used the projected rotational velocities from \citetalias{Nardini} (available for all stars except the SB2 systems) to estimate the rotation frequency range, assuming $\sin i$ between 0.1 and 1. The rotation frequency in d$^{-1}$ is given by:
\begin{equation}
    f_{\rm rot} \approx 0.02 \, {\rm v_{\rm rot}}\,({\rm R_{\odot}}/R),
    \label{eq:nu_rot}
\end{equation}
where $\rm v_{\rm rot}$ is the equatorial rotational velocity in units of km\,s$^{-1}$ and $R$ is the stellar radius in units of solar radii.
Estimates for the stellar radii were adopted from the tabulated values determined with different methods (eclipsing binaries and non-radial oscillations) by \citet{Gray_2021} and are reported in Table~\ref{tab:approx radii}. 

For the giant stars, we adopted the same radii as those used for dwarfs of the corresponding spectral types, as no reliable empirical or theoretical radius calibrations are available specifically for evolved stars. This approximation is justified by the fact that, within a given spectral type, the radius variation between the zero-age main sequence (ZAMS) and the TAMS is significantly smaller (i.e. it increases by a factor of about 2 for a 7~M$_{\odot}$ star) than the propagated uncertainties introduced by the unknown inclination and the measured projected rotational velocity, which are of the order of $\pm30$~\kms.
For example, we examined the difference in radius between the ZAMS and TAMS using the grid of MESA models from \citet{Johnston_2019a}. We considered B1- and B2-type stars (i.e.\ initial masses of 12 and 7 M$_\odot$, respectively), and used the lowest and the highest values of the CBM (i.e.\ 0.005 and 0.040) available in the grid. The stellar radius increases by approximately a factor of two between the ZAMS and the TAMS for the lowest CBM efficiency, and up to a factor of three for the highest CBM efficiency case.
Given that the spectral type classification does not easily distinguish stars on the main-sequence, from those in the sub-giant or giant evolutionary stages, and that the amount of CBM for each stars is unknown, this level of uncertainty is acceptable for the purpose of this work.

\begin{table}
\centering
\caption{Approximate radii for dwarf B-type stars from \citet{Gray_2021}.}
\begin{tabular}{cc}
\hline \hline
Spectral Type & R [R$_{\odot}$]  \\ \hline
B0  & 6.50  \\
B1  & 4.75  \\
B2  & 4.20  \\
B3  & 3.50  \\
B5  & 2.69  \\
B7  & 2.44  \\
B9  & 2.07 \\
\hline
\end{tabular}
\label{tab:approx radii}
\end{table}

\subsection{Frequency analysis}\label{sec:freq analysis}

For frequency analysis we used TESS sectors with a 200-sec cadence, which are those from the second extended mission, unless two or more consecutive sectors were available in earlier sectors. For example, stars in NGC~1960 had three consecutive sectors with a 10-min cadence (i.e. sectors 43, 44 and 45). 

For isolated stars (i.e.\ those with $C_{\rm ratio} \leq 1$) and identified as $\beta$~Cep, hybrids, and SPB stars, we identify significant frequencies using standard iterative pre-whitening methods \citep[see][]{Papics2012, BowmanMichielsen2021} using the longest uninterrupted chain of TESS sectors of the extracted and reduced light curves (see Section~\ref{sec:lc extraction}). The iterative prewhitening is achieved using the \textsc{period04} software \citep[][]{period04}. Following \cite{Baran_2021}, we adopt a significance threshold of S/N $\geq 4.6$ in the amplitude spectrum, which corresponds to a false-alarm probability of approximately 0.1~per~cent. This is better suited for space-based observations, since the large number of data points and longer time spans compared to typical ground-based data sets require a higher S/N threshold. Moreover, we tested using a lower S/N threshold of four but it did not significantly improve the number of significant peaks detected. 
The noise is calculated from a frequency window of width 2~d$^{-1}$ centred on the extracted frequency. 
A smaller window size (i.e. $<1$~d$^{-1}$) is not ideal for the high density of frequencies present in the amplitude spectra of fast rotators. On the other hand, a larger window size may include additional pulsation frequencies. This means the choice of window size is always a compromise of different effects.

Following extraction, we checked for unresolved frequencies in the frequency lists of all stars. According to the Loumos–Deeming resolution criterion \citep{Loumos1978}, two frequencies are resolved if:
\begin{equation}\label{eq:loumos-deeming}
    |f_{j} - f_{k}| \geq \frac{2.5}{\Delta T} ~ ,
\end{equation}

\noindent where $f_i$ and $f_j$ are a pair of significant frequencies, and $\Delta T$ is the time span of the light curve. In instances where the criterion was not satisfied, the lower-amplitude frequency was discarded.

The next step was the identification of harmonics and combination frequencies. A frequency, $f_i$, is classed as a harmonic or a combination if:
\begin{equation}\label{eq:comb frequencies}
    |f_{i} - (mf_{j} - nf_{k})| \leq \frac{2.5}{\Delta T} ~ ,
\end{equation}

\noindent where $f_{j}$ and $f_{k}$ are the parent frequencies and $\{n,m\} \in \{0,\pm1,\pm2,\pm3\}$ (see \citealt{Papics2012, Bowman2017}). 
The two highest-amplitude frequencies are considered parent frequencies and assumed to be self-excited pulsation mode frequencies \citep{Kurtz_2015}, and the combination frequencies are then determined using Eq.~(\ref{eq:comb frequencies}). 
Since harmonics and combination frequencies may not represent independent self-excited pulsation modes and are generally not used in asteroseismic modelling (see discussion by \citealt{Bowman2017}), they are not included in the final list of significant and independent pulsation frequencies for each star.

Appendix Table~\ref{tab:significant_frequencies} reports the significant independent frequencies along with their amplitudes and phases for each pulsating star. Interesting pulsating stars are discussed in Section~\ref{sec:variability in clusters}.

\subsection{\textsc{tlusty} versus Gaia ESP-HS atmospheric parameters}\label{sec:tlustyvsgaia}

In \citetalias{Nardini}, the atmospheric parameters were determined for apparently single stars and SB1 systems using a grid of rotationally broadened \textsc{tlusty} models \citep[][]{tlusty_1995, tlusty_2007}. Using a similar approach to that used by \citet{bodensteiner_vsini}, a grid-based search of BSTAR2006 \textsc{tlusty} models and a $\chi^2$ minimisation, the best-fitting projected rotational velocity was determined for each star. Only specific lines were used in the fitting process, especially for Be stars. For example, for Be stars only the wings of the Balmer lines and metal lines were used in the fit, since these were least affected by emission.
In \citetalias{Nardini}, the distributions of projected rotational velocity were found to be statistically consistent across the four clusters.

In this work, we compare the values determined using the \textsc{tlusty} grid in \citetalias{Nardini} with those derived by the Extended Stellar Parametrizer for Hot Stars \citep[ESP-HS;][]{Fouesneau_2023, Fremat_esphs} catalogue from data release 3 (DR3) of ESA's Gaia mission \citep{Gaia1, GaiaDR1, GaiaDR2, GaiaDR3}.
The ESP-HS module is applied to O-, B-, and A-type stars only after the source has been identified as a hot star by the Extended Stellar Parametrizer for Emission-Line Stars (ESP-ELS), which classifies both spectral type and the presence of emission lines. The ESP-HS module uses Gaia's BP/RP low-resolution spectra and, when available, the medium-resolution RV spectrometer (RVS) data. Stellar parameters are inferred by minimising the weighted squared residuals against synthetic spectral libraries. When RVS spectra of sufficient quality are available, their weighted squared residuals are included in the fit \citep{Fouesneau_2023, Fremat_esphs}.

\citet{Fouesneau_2023} compared the astrophysical parameters obtained with the ESP-HS pipeline with several previous catalogues of early-type stars. They found good agreement in $T_{\rm eff}$ below 25\,000~K, but a systematic underestimation of the Gaia values compared to spectroscopy above this temperature. Likewise, the scatter in the $\log\,g$ differences between Gaia and the reference catalogues increases from about 0.2~dex for A-type stars to roughly 0.4~dex for O-type stars. Similarly, \citet{Shi_2023_esphs} report that the Gaia $T_{\rm eff}$ values obtained by ESP-HS are reliable for pulsating hot main-sequence stars with $T_{\rm eff} \lesssim 30\,000$~K from a comparison of the ESP-HS derived parameters to those from low-resolution LAMOST spectra \citep{Guo_2021_lamost}. However, the authors report only the effective temperature and mention that the consistency between the LAMOST stellar parameters (likely meaning also the surface gravity and the projected rotational velocity) and the ESP-HS parameters is poor, and therefore, this is why they are not mentioned in their paper explicitly.

The ESP-HS module of the Gaia mission has several limitations. For example, not all Be stars are correctly identified by ESP-ELS, meaning those misclassified as non-emission-line stars have unreliable stellar parameters. Of the 15 Be stars in our sample identified in \citetalias{Nardini}, only six are correctly classified as emission-line stars by ESP-ELS. Moreover, a similar problem arises for SB2 systems in that the spectral resolution of Gaia's RVS instrument is insufficient to detect two sets of spectral lines in most SB2 systems. Yet, since the flux of both components is present in the Gaia spectra, the derived atmospheric parameters are typically not reliable. Additionally, it is worth noting that stellar parameters from the Gaia ESP-HS catalogue are not available for five stars in our sample.

\begin{figure}
    \centering
    \includegraphics[width=0.99\linewidth]{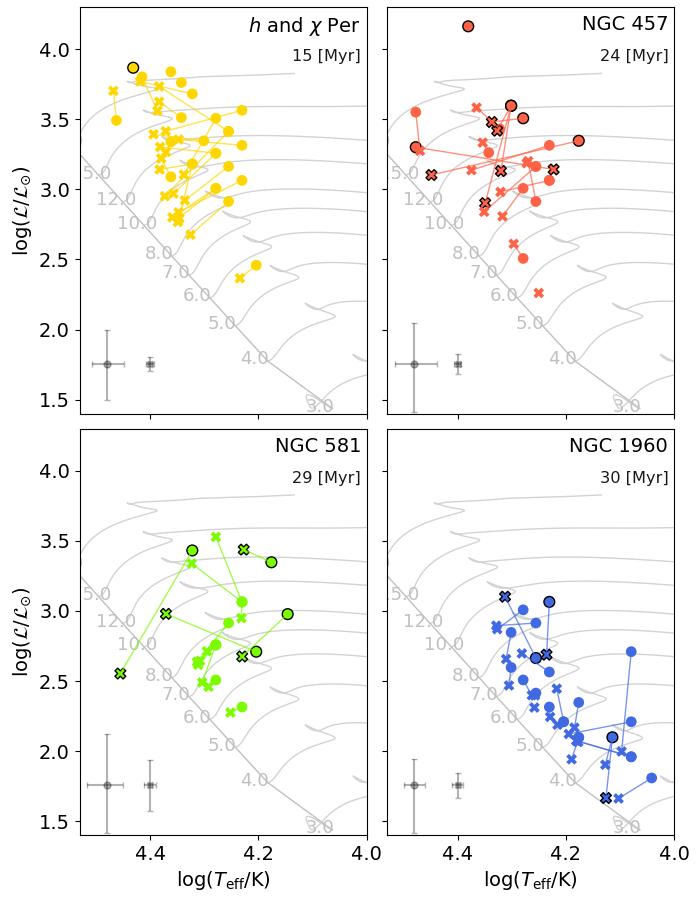}
    \caption{Spectroscopic HR~diagrams of the four clusters. Circles indicate the stellar parameters obtained using the \textsc{tlusty} atmospheric models by \citetalias{Nardini}, while crosses indicate values from the Gaia ESP-HS catalogue. The Be stars are denoted with a black edge colour. The grey lines are stellar evolutionary tracks, which are labelled in units of M$_{\odot}$ and were computed by \citet{Burssens_2020}.
    The size of typical error bars for each cluster is shown in the bottom left, both for the \textsc{tlusty} (left) and for the ESP-HS (right). The ages below the name of each cluster are from \citet{Dias2021}.}
    \label{fig:tlusty_gaia}
\end{figure}

In Fig.~\ref{fig:tlusty_gaia} we show the spectroscopic HR~diagrams of the four clusters, which use spectroscopic luminosity defined as $\mathcal{L} = T_{\rm eff}^4 / g$ \citep[][]{Langer_sHRD} as the ordinate axis. We compare the stellar parameters derived from the $\chi^2$ minimisation procedure using {\sc tlusty} models with those obtained by the ESP-HS module from the Gaia mission.
The older cluster NGC~1960 \citep[30~Myr; ][]{Dias2021} is dominated by late-B stars, and we find a smaller disagreement between our {\sc tlusty} parameters and those from Gaia. The average absolute difference for the effective temperature is $\sim2000$~K, while for the surface gravity is $\sim0.25$~dex.
The largest discrepancy appears in $h$ and $\chi$ Per, which is the youngest cluster \citep[ NGC~869 and NGC~884; 15~Myr; ][]{Dias2021} and hosts the most massive stars in our sample. For example, our {\sc tlusty} results tend to place these stars beyond the terminal-age main sequence (TAMS) in Fig.~\ref{fig:tlusty_gaia}, which is likely because the  {\sc tlusty} approach has underestimated the $\log\,g$ values. This thus causes the spectroscopic luminosity to be overestimated. The average absolute differences are also larger compared to the older cluster: while the difference in $T_{\rm eff}$ is $\sim3500$~K, the difference in $\log\,g$ is $\sim0.4$~dex.

All Be stars in our sample show a particularly substantial discrepancy in their {\sc tlusty} and Gaia parameters. This is expected because most of them are misclassified as non-emission-line objects by the Gaia mission and have subsequently been treated as non-emission-line B stars. 
For this reason, the parameters determined by the \textsc{tlusty} grid were used for the Be stars as well as for the five stars that do not have ESP-HS parameters, while for the remaining of the sample we used the ESP-HS parameters in Fig.~\ref{fig:HRD}. On the other hand, our {\sc tlusty} analysis of Be stars suffers from limitations of neglecting the effects of fast rotation and time-dependent emission features. Therefore, it is not surprising that the parameters of such stars have large uncertainties and a large discrepancy with the Gaia parameters.

\section{Variability in the four open clusters}\label{sec:variability in clusters}

We use the TESS light curves and frequency spectra to classify the photometric variability of the 63 stars out of 82 with an acceptable level of contamination in our sample across the four clusters.
The results of the variability classification are reported in the last column of Table~\ref{tab:stars_list}. 
Among the 63 stars, 29 are $\beta$~Cephei stars, two are SPBs, 14 are pulsating Be stars, seven are hybrids, five are photometric binaries, three are rotational modulation stars, one is a $\delta$ Scuti star, and two are constant. 
Based on the sample of 82 stars, we find a photometric variability fraction of $77\pm5$~per~cent, but if only the stars not in crowded regions are considered (63), we obtain a variability fraction of $97\pm2$~per~cent and a pulsator fraction of $84\pm5$~per~cent. 
It is well known that the majority of low-mass stars in older clusters pulsate, both during and beyond the MS \citep[see ][]{Stello_2010, Corsaro_2012, Sandquist_2020, Li_2024a, Li_2026, Fritzewski_2026}. This work demonstrates a high incidence of pulsations for more massive early-type MS stars in younger open clusters, expanding on previous work for other studies of young open clusters \citep[e.g.][]{Saesen_2010, Sharma_2022}.

\subsection{Pulsating Be stars}\label{sec:be stars}

\begin{figure*}
    \centering
    \includegraphics[width=0.85\linewidth]{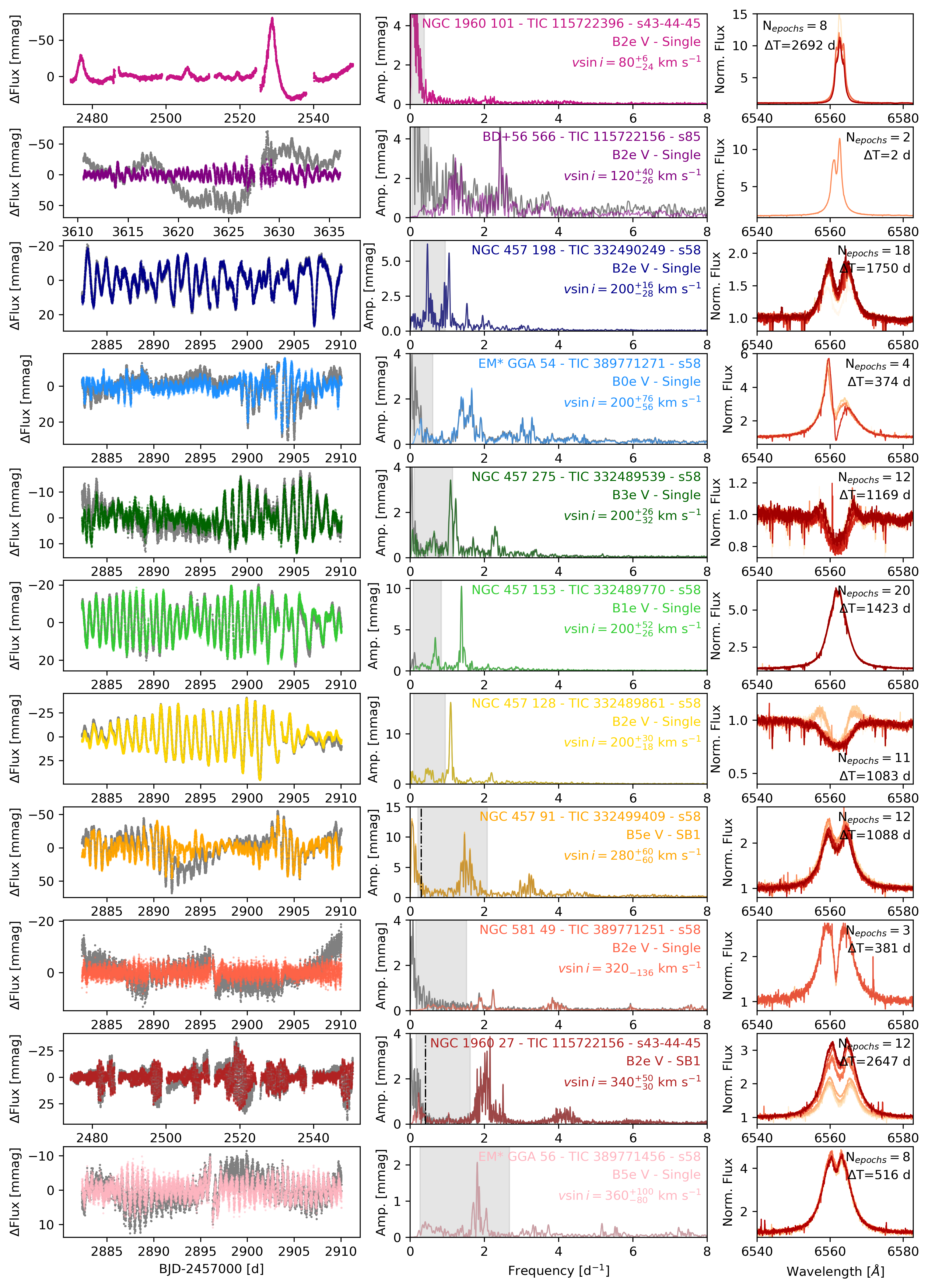}
    \caption{Light curves (left panels), frequency spectra (central panels), and available {\sc Hermes} spectra of the \ion{H$\alpha$}{} line (right panels) for the 11 non-contaminated pulsating Be stars across the four clusters. The stars are ordered top-to-bottom in increasing projected rotational velocity. In the left panels, we show both the non-detrended (light grey curve) and the detrended light curves (in colour), and the corresponding frequency spectra from these two light curves are shown in the middle panels in the same colour. The grey shaded regions in the frequency spectra represent the frequency range where rotational modulation may be expected given the measured $v\,\sin\,i$ and estimated radius of each star. The black vertical dash-dot lines in the middle panels show the orbital frequency for identified SB1 systems. In the right panels, we also report the number of spectroscopic epochs available and the total time coverage of the {\sc Hermes} spectra ($\Delta$T). The spectra are ordered from lighter (earliest epoch) to darker (latest epoch) colours. We note that the spectra are not contemporaneous with the TESS sectors. 
    }
    \label{fig:green_be}
\end{figure*}

The photometric and spectroscopic variability characteristics of Be stars known in the literature are detected in all of the Be stars analysed in this work. These include complex beating patterns in their light curves and multi-periodic non-radial gravity~modes, which can form frequency groups due to the harmonics and combination frequencies \citep[][]{Kurtz_2015}. These features are common but not exclusive to Be stars, and are also generally found in rapidly rotating stars \citep[][]{Saio_2013, Kurtz_2015, Naze_2024}.

In \citetalias{Nardini}, a total of 15 Be stars were identified in multi-epoch {\sc Hermes} spectroscopy. In this work, 14 of them show pulsations in their TESS light curves. Only one Be star, V$^{*}$~V1122~Cas, is too contaminated to determine whether the star is pulsating or if a nearby contaminating star is instead. Specifically, there is a variable blue supergiant in the same TESS FFI pixel of V$^{*}$~V1122~Cas, whose SLF variability dominates the variability of the Be star.
Three Be stars (NGC~457~6, NGC~457~14, NGC~1960~47) are tentatively classified as pulsators, because there is considerable sector-to-sector variability in their light curves, indicating they are also likely contaminated. This means we cannot reliably perform any frequency analysis since it is not clear if the changes are due to stellar variability, contamination or instrumental effects.
For the remaining 11 Be stars, the amount of contamination was low so it was possible to extract TESS light curves, calculate frequency spectra and perform iterative pre-whitening. Figure~\ref{fig:green_be} shows the light curves, the frequency spectra, and the {\sc Hermes} spectra zoomed-in on \ion{H$\alpha$}{} for these 11 pulsating Be stars. The grey shaded region in each of the central panels marks the estimated frequency range in which one may expect rotational modulation, which is calculated using Eq.~(\ref{eq:nu_rot}).

We detect multi-periodic gravity-mode pulsations in all 11 non-contaminated Be stars across the four clusters. For all of them, the peak with the highest amplitude has a frequency below 3~d$^{-1}$. Almost all of the 11 stars show two or more frequency groups in their frequency spectra, which are consistent in frequency across all available sectors but vary expectedly in amplitude.
Frequency groups extending beyond 4~d$^{-1}$ are only observed for the faster rotating Be stars ($v\,\sin\,i > 200$~km\,s$^{-1}$; shown in the bottom four rows of Fig.~\ref{fig:green_be}). 
This may lead to an incorrect classification of these four stars (NGC~457~91, NGC~581~49, NGC~1960~27, and EM$^*$~GGA~56) as $\beta$~Cep stars because of the higher pulsation frequencies, but their rapid rotation can shift the gravity modes to higher frequencies \citep[][]{Labadie_Bartz_2022, Aerts_2024_fastrot}.

In terms of binarity, Be stars generally show a lack of main-sequence companions \citep[][]{Bodensteiner_2020_lackbe}, so it is not surprising that in this work only two Be stars are candidate binary systems (NGC~457~91 and NGC~1960~27). However, this does not exclude the possibility of companions being either a stripped star (see e.g. \citealt{Abdul-Masih_2020, shenarLB1, Bodensteiner_hr6819}) or a compact object \citep[][]{Janssens_2025}. The presence of a companion might influence the decretion disc of a Be star, and cause the so-called violet-to-red (V/R) cycles attributed to density waves propagating through the disc \citep[][]{Rivinius_2016}.  
The binarity status of NGC~457~91 and NGC~1960~27 is discussed in more detail in Section~\ref{sec:binary systems}.

Making a comparison or ensemble analysis is challenging since the Be phenomenon is transient and the stars in our sample show a different range of behaviours. Each individual pulsating Be star is discussed in Appendix~\ref{appendix:individual_Be}.

\subsection{$\beta$~Cep stars, SPB stars, and hybrids}

\begin{figure*}
    \centering
    \includegraphics[width=0.90\linewidth]{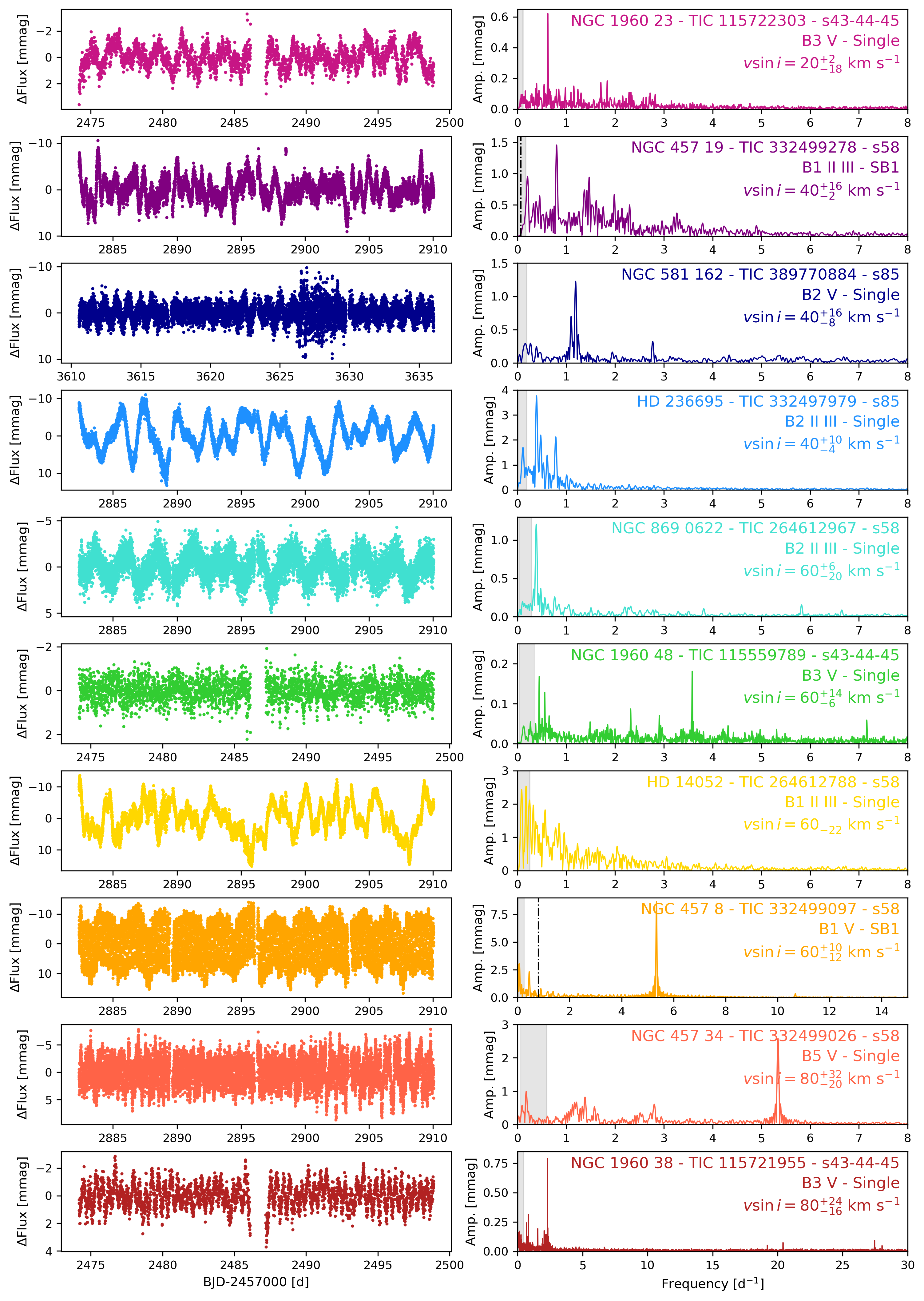}
    \caption{Light curves and frequency spectra for the slower rotating $\beta$ Cep and hybrid stars that are not contaminated. The layout and styling are the same as Fig.~\ref{fig:green_be}, except the \ion{H$\alpha$}{} panel is excluded since there are no emission features.     }
    \label{fig:betacep_lowvsini}
\end{figure*}

\begin{figure*}
    \centering
    \includegraphics[width=0.90\linewidth]{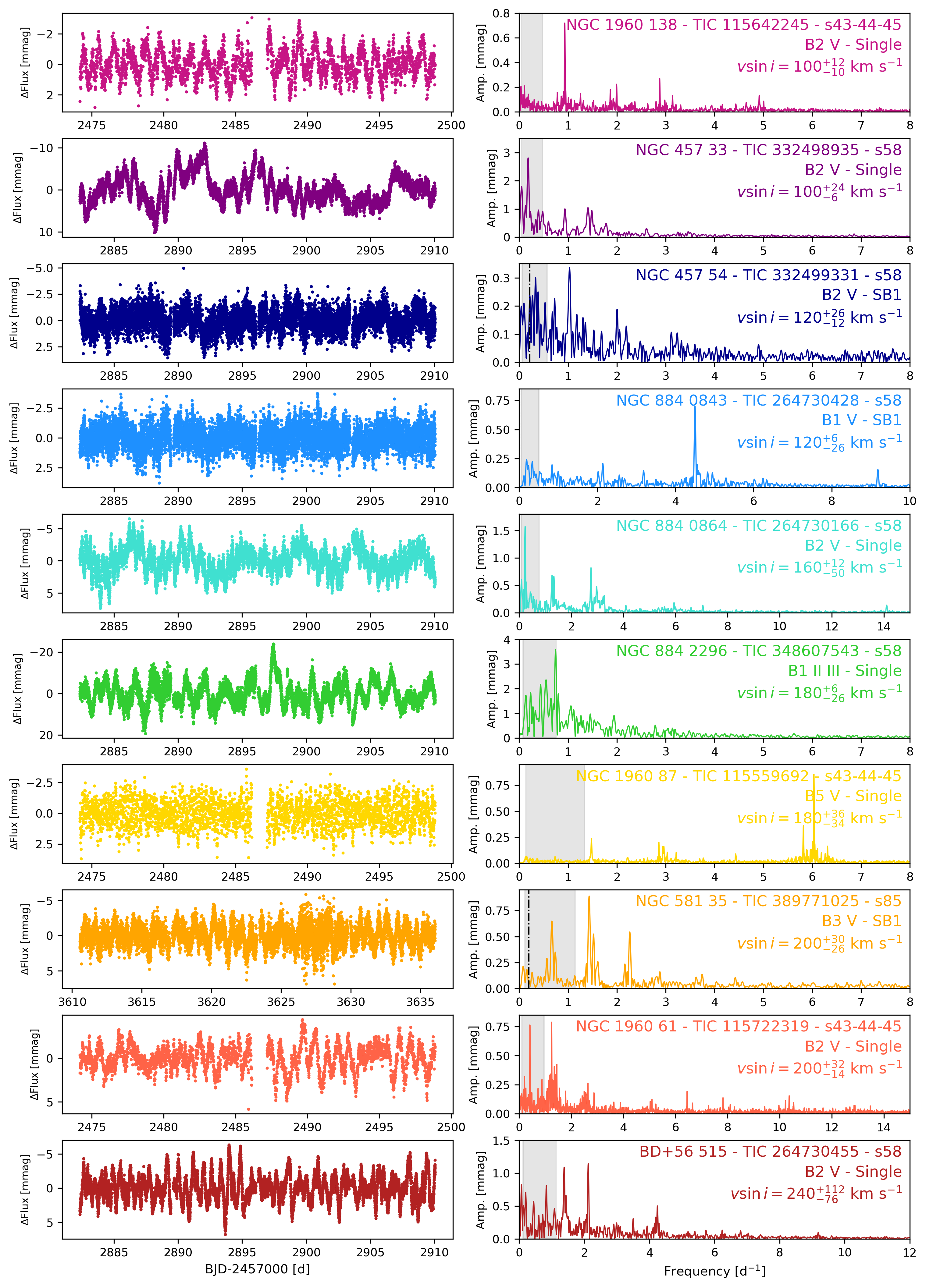}
    \caption{Same as Fig.~\ref{fig:betacep_lowvsini} but for the $\beta$~Cep and hybrid stars with v$\sin i\geq100$~km\,s$^{-1}$.}
    \label{fig:betacep_highvsini}
\end{figure*}

\begin{figure*}
    \centering
    \includegraphics[width=0.90\linewidth]{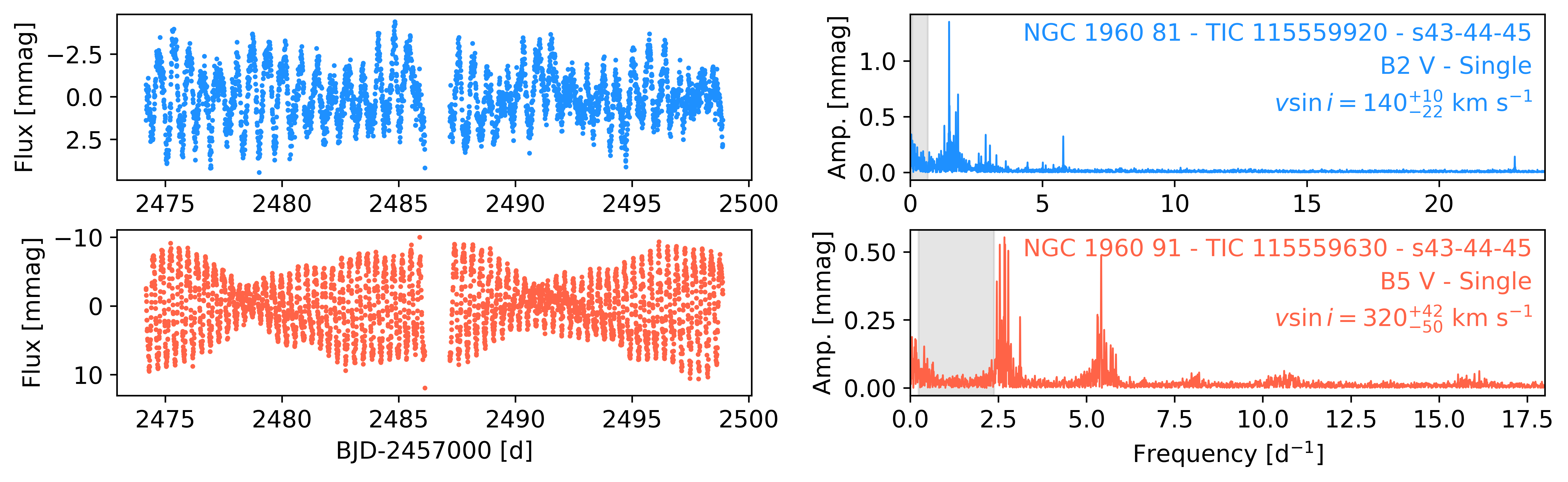}
    \caption{Same as Fig.~\ref{fig:betacep_lowvsini} but for the SPB stars.}
    \label{fig:spb_lc}
\end{figure*}

Traditionally, $\beta$~Cep stars have spectral types from O9 to B2 on the main sequence \citep{Stankov_2005}, corresponding to masses of about 8 to 30~M$_{\odot}$. However, modern space photometry has revealed that the low-radial order pressure and gravity modes of $\beta$~Cep stars are found among both higher and slightly lower masses \citep{Burssens_2020, Fritzewski_2025, Thomson-Paressant_2026}. The pressure-mode instability region from \citet{Burssens_2020} has a minimum mass of about 8~M$_{\odot}$ and yet, although rare, some $\beta$~Cep stars have been found to have masses as low as 5-6~M$_{\odot}$ \citep[e.g.][]{Burssens2019}. In our sample, we find $\beta$~Cep stars spanning almost the full mass range, from about 5 to about 25~M$_{\odot}$ with the lowest-mass $\beta$~Cep stars being located outside the pressure-mode instability region, as shown in Fig.~\ref{fig:HRD}. For example, NGC~457~34, NGC~1960~87, and NGC~1960~91 have spectral types of B5 and atmospheric parameters that confirm them to be quite cool for $\beta$~Cep stars. 

On the other hand, distinguishing pressure modes from gravity modes in mid-B stars is challenging when relying solely on the observed frequency range. Some pulsators exhibit two distinct sets of self-excited pulsation spectra, where each set corresponds to modes typically associated with different established classes of variable stars. Such objects are therefore commonly referred to as hybrid pulsators \citep{Handler2009}.
For this reason, we classify $\beta$~Cephei stars as early-B stars showing predominantly high-frequency ($f \geq 4$ d$^{-1}$) pulsations, while stars of spectral type B5 or later exhibiting gravity modes are classified as SPB stars or, when both sets of frequencies are present, as hybrids.

Overall, we find that $\beta$ Cep stars are the most common class of pulsators in the sample, which is not unexpected considering it is dominated by early-B stars. However, two clear examples of SPB stars exist (NGC~1960~81 and NGC~1960~91), which have higher frequency gravity modes and frequency groups because of their rapid rotation.
Figures~\ref{fig:betacep_lowvsini} and \ref{fig:betacep_highvsini} show the TESS light curves and frequency spectra of the 20 non-contaminated $\beta$~Cep stars and hybrids, while Fig.~\ref{fig:spb_lc} shows the light curves and frequency spectra of the two SPB stars. 
We classify two stars as hybrids: NGC~457~34 and NGC~1960~87 on the basis that their pulsation frequencies span a wide range in frequency and their spectral type places them in the overlapping region for pressure and gravity modes (c.f. Fig.~\ref{fig:HRD}). 
The remaining 18 stars include six in $h$ and $\chi$ Persei (i.e. NGC~869 and NGC~884, respectively), five in NGC~457, two in NGC~581 and five in NGC~1960. The somewhat lower number of pulsators in NGC~581 is likely due to the high crowding of the field rather than an intrinsically lower pulsator fraction. 
For most stars with one TESS sector, we find significant frequencies up to 8~d$^{-1}$ with a few exceptions with higher frequencies (e.g. NGC~457~8, NGC~869~0843, NGC~884~0864, and BD+56~515). For stars in NGC~1960 and their longer TESS light curves of three consecutive sectors, we find significant frequencies up to about 30~d$^{-1}$ (e.g. NGC~1960~38).

The detection of rotationally split multiplets enables mode identification in terms of spherical harmonic geometry and facilitates forward asteroseismic modelling \citep[e.g.][]{Burssens_2023, Vanlaer_2025}. However, for moderate-to-fast rotators, asymmetric multiplets are expected because of centrifugal deformation \citep[][]{Aerts_2024_fastrot}. Almost all of our targets are fast rotators, so it is likely for this reason that we did not detect any symmetric rotational splitting for any of the $\beta$~Cep stars.
For the fastest rotators (i.e. $v\,\sin\,i \geq 200$~km~s$^{-1}$), frequency groups similar to those in pulsating Be stars are present. The SPB star NGC~1960~91 is perhaps the best example of this. It has a projected rotational velocity of 320~km~s$^{-1}$ and the TESS light curve shows complex beating patterns and frequency groups, as shown in Fig.~\ref{fig:spb_lc}, but it does not show any emission in the {\sc Hermes} spectra. Its frequency spectrum shows frequency groups above 4~d$^{-1}$, but upon inspection, they appear to be harmonics of the group with higher amplitudes at $\sim2.5$~d$^{-1}$.
Thus, this star could either be considered a rapidly rotating SPB star and not a Be star, or a hybrid star, and is an example of how classifying pulsators without mode identification is not straightforward.

\section{Revisiting spectroscopic binaries with TESS}\label{sec:binary systems}

In \citetalias{Nardini}, RV variability from multi-epoch {\sc Hermes} spectroscopy was used to identify spectroscopic binaries across the four young open clusters. To investigate the possible degeneracies between pulsations and binarity when identifying spectroscopic RV variability, we compare the dominant photometric period ($P_{\rm phot}$) identified in this work based on TESS light curves with the spectroscopic period ($P_{\rm spec}$) determined in \citetalias{Nardini}.
We note that the eight additional stars in this work that were added to the sample of 74 stars from \citetalias{Nardini} were previously excluded because either their spectral types were later than B5, or only one or two spectra were available, preventing an RV analysis. This means that these eight stars do not yet have any evidence for binarity.

\begin{figure}
    \centering
    \includegraphics[width=0.95\linewidth]{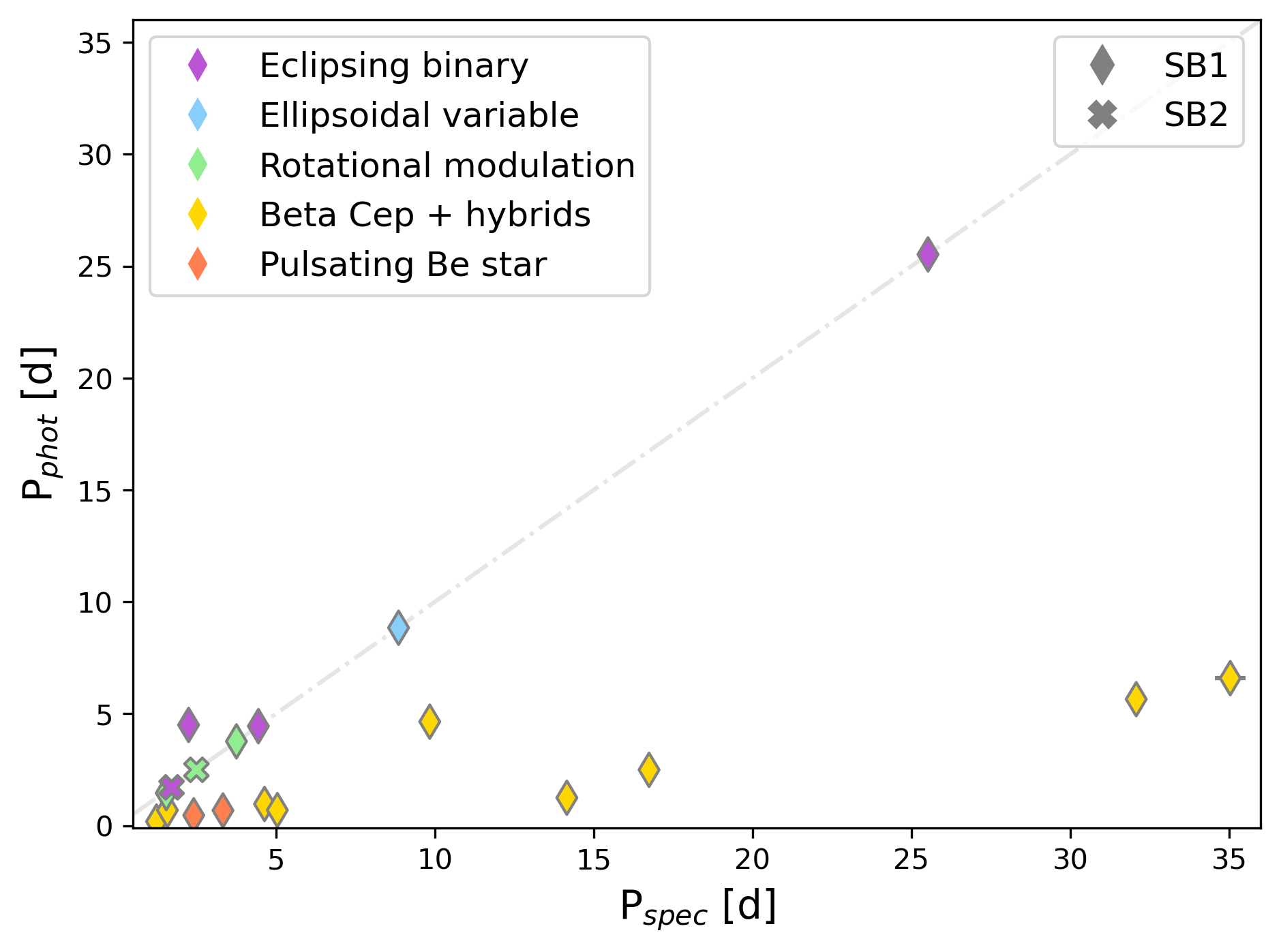}
    \caption{Dominant photometric period from the TESS light curve analysis in this work versus the spectroscopic period from the RV analysis for  the subset of 28 binaries reported in \citetalias{Nardini}. Symbol colours denote the different classes of pulsators, and different symbol styles denote the type of binaries (SB1 or SB2). The dashed grey line is the bisector. Error bars are included but they are generally smaller than the symbol size.    }
    \label{fig:1srharm_orbper}
\end{figure}

In Fig.~\ref{fig:1srharm_orbper}, we compare the dominant photometric period from our analysis of TESS light curves to the inferred spectroscopic binary period for the subset of 23 candidate binary systems. The remaining five stars that are not included in Fig.~\ref{fig:1srharm_orbper} either have too high contamination or did not show any significant photometric variability in their light curve.
If the spectroscopic period is consistent with the photometric one, all photometric binaries (i.e. EBs, ellipsoidal variables) and stars with rotational modulation should lie on the bisector. Only one of these is an anomaly, EM$^*$~GGA~52, which is an EB with a photometric period that is twice the reported spectroscopic period from \citetalias{Nardini}. Therefore, the previously reported binary period of $P_{\rm spec}=2.2507 \pm 0.0002$~d is an alias of the true binary period of $P_{\rm phot}=4.502164 \pm 0.000003$~d.

All other stars that do not lie on the bisector in Fig.~\ref{fig:1srharm_orbper} have $P_{\rm phot}$ that significantly differs from their reported $P_{\rm spec}$. All such systems are pulsators (i.e. $\beta$~Cep, hybrid, or Be stars). Thus, it is likely that the binary period inferred from the spectroscopic RV analysis is correct, and that the dominant photometric period is due to pulsations.

Closer inspection of the Be stars in Fig.~\ref{fig:1srharm_orbper} shows small values for both $P_{\rm spec}$ and $P_{\rm phot}$. As mentioned in Section~\ref{sec:be stars}, it was not possible to extract the light curves of all Be stars in our sample due to contamination. Of the four Be stars in binaries reported by \citetalias{Nardini}, only two did not show a high level of contamination, NGC~457~91 and NGC~1960~27, which are included in Fig.~\ref{fig:1srharm_orbper}. With constraints on the pulsation periods now known from the TESS light curves, it becomes clear that the previously identified value of $P_{\rm spec}$ is likely a pulsation period and not a binary period. 
Moreover, the orbital solutions reported by \citetalias{Nardini} for the two purported Be-star binary systems were not robust (see Appendix of \citetalias{Nardini}), and the RV variability was also quite high (i.e. $\Delta{\rm RV}= 41 \pm 8$~km\,s$^{-1}$ for NGC~457~91 and $\Delta{\rm RV}= 29 \pm 8$~km\,s$^{-1}$ for NGC~1960~27). With our new analysis of TESS light curves and the complementary {\sc Hermes} spectroscopy, we conclude that these two Be stars are apparently single stars, with RV variability caused solely by pulsations. This suggests that the 20~km\,s$^{-1}$ threshold commonly used for distinguishing binarity from pulsations and other variability effects in massive stars \citep[][]{banyard, Bodensteiner_2025_bloem, Villasenor_2025} may need to be larger for Be stars (see also \citealt{Thomson-Paressant_2026}). The binary status of the other two Be stars that were also classified as binaries but with small values of $P_{\rm spec}$ should also be questioned, but since it was we could not extract the TESS light curves, it is not possible to make a final statement if the determined orbital period is actually their pulsation period.

\begin{figure}
    \centering
    \includegraphics[width=0.95\linewidth]{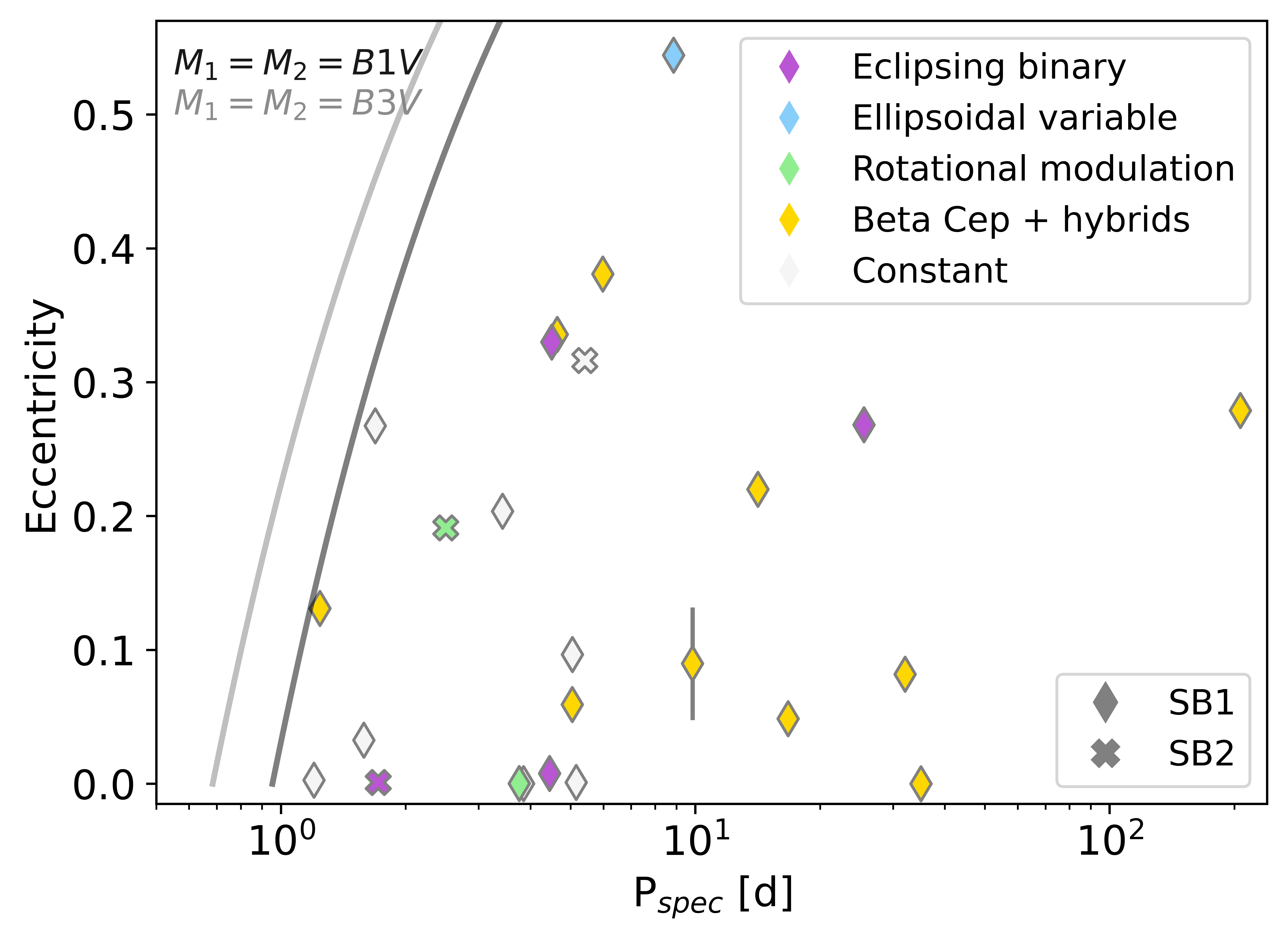}
    \caption{Eccentricity as a function of period for the revised subsample of 25 binaries. We use the same colours and symbol styles as Fig.~\ref{fig:1srharm_orbper} for the different classes of pulsators and types of binaries. The two solid lines correspond to the period and eccentricity a system should have for the semi-major axis to be equal to the sum of the radii of the two components. The darker purple colour is for a system composed of two B1 V stars and the lighter colour for two B3 V stars. Error bars are included but they are generally smaller than the symbol size.    }
    \label{fig:new_p_vs_e}
\end{figure}

If these Be stars in short-period binaries are reclassified as single stars, we can re-calculate the observed binary fraction of the two clusters as $53 \pm 11$~per~cent for NGC~457, and $27\pm 11$~per~cent for NGC~1960. Note that these values are computed based on the sample of 18 stars analysed in \citetalias{Nardini}, thus the five later type stars are similarly excluded for consistency reasons. These observed binary fractions are 5 and 3~per~cent lower, respectively, compared to those derived in \citetalias{Nardini}. Using the same methodology as \citetalias{Nardini}, the bias-corrected intrinsic binary fractions are reduced by 8 and 13~per~cent, respectively, and are $75^{+28}
_{-26}$~per~cent for NGC~457 and $68^{+32}_{-43}$~per~cent for NGC~1960. Although slightly lower, these intrinsic binary fractions are still consistent with each other within their 1$\sigma$ confidence intervals, and the other two clusters studied in this work. Therefore, the conclusion of \citetalias{Nardini} is still valid that the majority of early B-type stars in galactic clusters of ages between 15 and 30~Myr are in binaries. 
In Fig.~\ref{fig:new_p_vs_e} we show the updated version of the eccentricity versus period diagram for the revised sample of 25 identified binaries (see Figure~5 of \citetalias{Nardini}), which is colour coded with their photometric variability class.

\subsection{Photometric binaries}
\begin{figure*}
    \centering
    \includegraphics[width=0.99\linewidth]{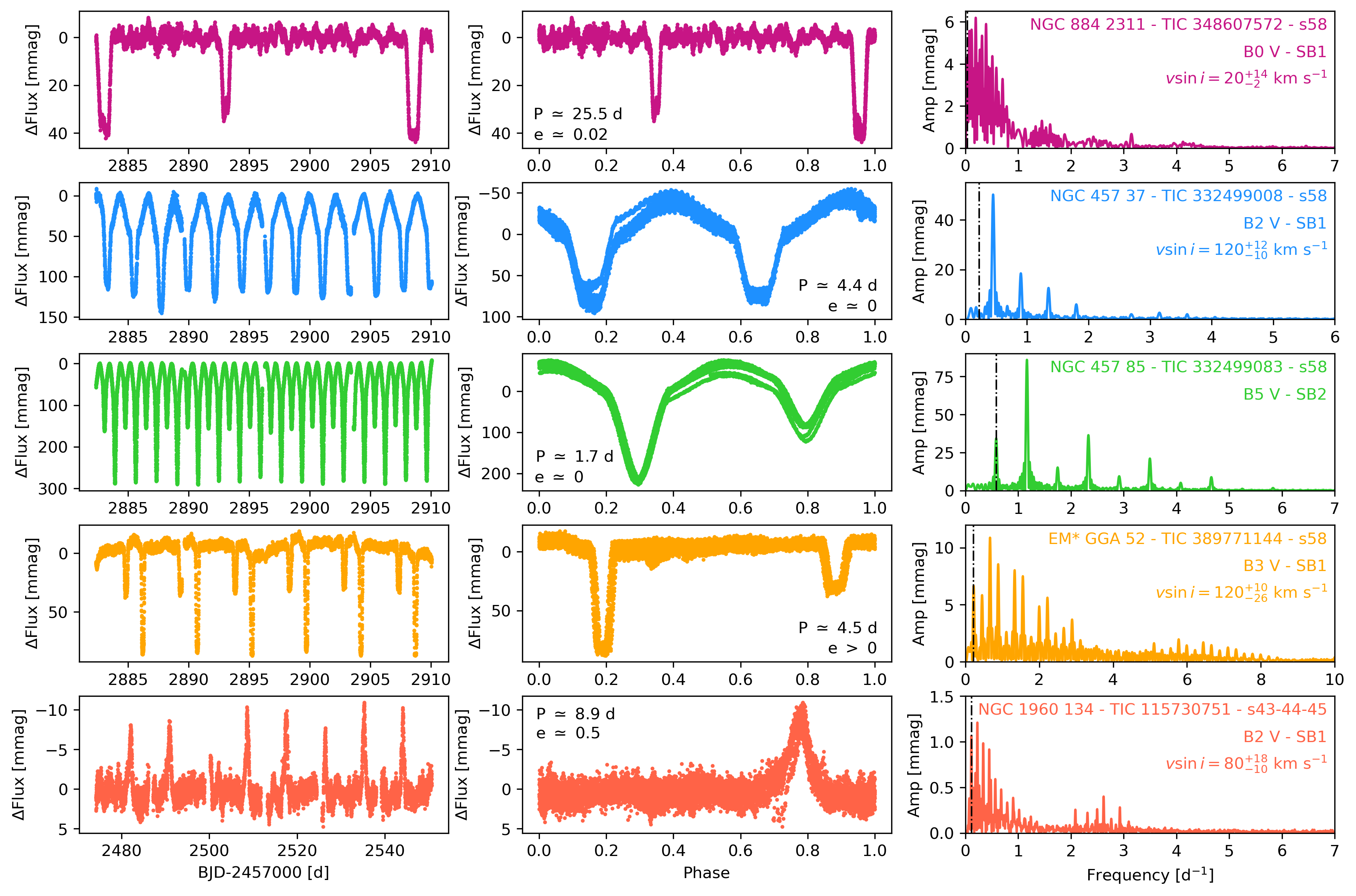}
    \caption{TESS light curves (left panels), phase-folded light curves (central panels), and frequency spectra (right panels) of the five photometric binaries identified from this work. The vertical dash-dot black lines in the third-from-the-right panels denote the orbital frequencies determined from spectroscopy in \citetalias{Nardini}.}
    \label{fig:green_eb}
\end{figure*}

We identify five photometric binaries in our sample. Their TESS light curves, phase-folded light curves, and frequency spectra are shown in Fig.~\ref{fig:green_eb}. Four of these are EBs, NGC~884~2311, NGC~457~37, NGC~457~85, and EM$^{*}$~GGA~52, and one is an ellipsoidal variable: NGC~1960~134.
The four EBs were all previously known, but the ellipsoidal variable is a newly detected system. The five systems were also identified as spectroscopic binaries in \citetalias{Nardini}, where the orbital solution was also determined and consistent with the photometric period except for EM$^{*}$~GGA~52 as mentioned previously, in which the spectroscopic period was an alias of the true binary period of $4.502164 \pm 0.000003$~d. Discussion of individual systems is provided in Appendix~\ref{appendix:individual_binaries}.

\subsection{Rotational modulation}
We also identify stars with rotational modulation in their TESS light curves, which is commonly associated with the presence of chemical abundance anomalies \citep{Stibbs1950, Bowman2018}.  
The three stars with TESS light curves dominated by rotational modulation are NGC~1960~8, NGC~1960~16, and NGC~884~2185. Their light curves, frequency spectra and spectroscopic variability of the \ion{He}{I} $\lambda$5875 line are shown in Fig.~\ref{fig:rot_modulation}. We discuss the three stars in Appendix~\ref{appendix:individual_rot_mod}.

\begin{figure*}
    \centering
    \includegraphics[width=0.99\linewidth]{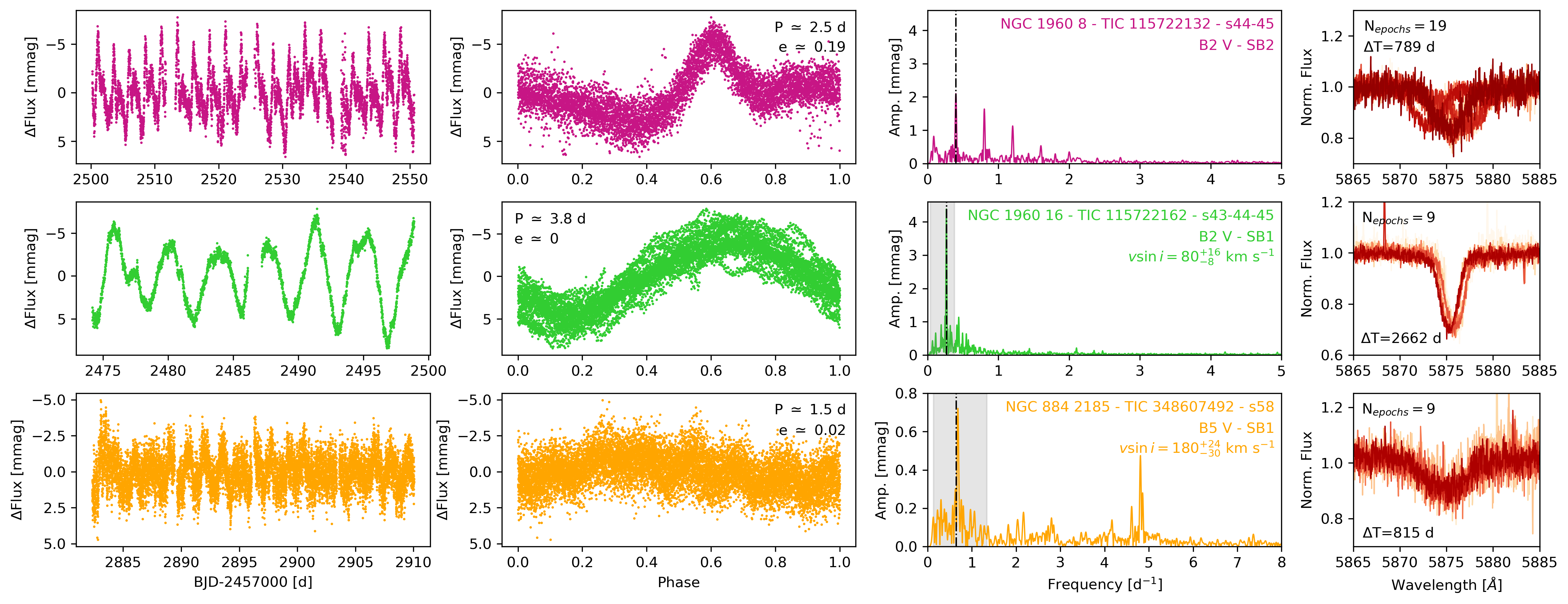}
    \caption{From left to right, the panels show the TESS light curves, the phase-folded light curves, and the frequency spectra of the three stars dominated by rotational modulation. The vertical dash-dot black lines in the right panels denote the purported orbital frequencies determined from spectroscopy in \citetalias{Nardini}, but here are demonstrated to be rotational modulation rather than binarity. In the right-most panel, the profile of \ion{He}{I} from multi-epoch {\sc Hermes} spectra is shown.}
    \label{fig:rot_modulation}
\end{figure*}

\section{Conclusions}\label{sec:conclusion}
In this second paper in a series focussing on massive stars in four young Galactic open clusters, we performed a photometric characterisation of 82 B-type stars with literature ages between 15 and 30~Myr. Using data from the NASA TESS space mission, we pushed the limits of photometric analysis in crowded cluster fields by applying different state-of-the-art light-curve extraction techniques. This allowed us to obtain reliable light curves for 63 stars and classify the variability of 80 per cent of the full sample. Among the stars with robust TESS light curves, $84 \pm 5$ per cent exhibit variability consistent with pulsations, which demonstrates that pulsations are common among B-type stars.

Excluding two constant stars, we classified the remaining 61 stars into three main categories: pulsating Be stars, coherent pulsators (including $\beta$ Cep stars, SPB stars, and hybrids), and photometric binaries (including eclipsing binaries and ellipsoidal variables), and stars showing rotational modulation. We find that all Be stars in the sample show pulsational variability, confirming the results of previous surveys. Their frequency spectra are dominated by pulsations in the gravity-mode regime ( i.e. below about 4~d$^{-1}$), and all exhibit the characteristic frequency groups commonly observed in this class of stars. 
$\beta$ Cep stars constitute the most common class of pulsators in the sample, which is expected given that B2 dwarfs dominate the sample. Low-radial order pressure modes are also detected in stars of later spectral type, suggesting that the $\beta$ Cep instability strip may extend to lower masses than predicted by instability strips. Rapidly rotating stars ($v \sin i \geq 200$~km~s$^{-1}$) show frequency groups caused by complex beating patterns in their light curves, similar to those observed in pulsating Be stars.
Correctly identifying SPB stars was challenging because most of the pulsators exhibit significant peaks above 4~d$^{-1}$, which is the oft-used threshold used to distinguish between pressure and gravity modes. However, we identify two stars all in the cluster NGC~1960 as SPB stars, and the analysis of their period spacing patterns is the subject of future work (Nardini et al., in prep). 

Finally, of the 30 spectroscopic binaries identified in \citetalias{Nardini}, eight of them were identified as photometric binaries in this work. These include four EBs, one ellipsoidal variable, but there are also three stars that show rotational modulation. The four EBs were all previously known in the literature, while the ellipsoidal variable and the stars with rotational modulation are newly identified. Two of the stars that show rotational modulation show undeniable signs of binarity in their spectra, with one of the two being an SB2 and the other having peak-to-peak RV variability higher than 40~km~s$^{-1}$, which is unlikely to be caused by pulsations. The third star does not show any significant RV variability in its spectra due to low S/N, thus there could be a bias in the binary detection. 
No new binaries were identified through photometry, and three systems (the aforementioned rotational modulation star and two Be stars) are deemed to be incorrectly classified as binaries in \cite{Nardini}. Thus, we conclude that the 20~km~s$^{-1}$ spectroscopic RV threshold often used to identify binaries should not be applied universally, especially not for Be stars.

Overall, this work highlights the importance of performing a multifaceted characterisation of massive stars using both spectroscopy and photometry, because both binarity and pulsations are almost ubiquitous and influence each other. In this work, we have identified interesting systems (e.g. EBs and pulsators) that warrant follow-up modelling, especially forward asteroseismic modelling. This demonstrates the power of the TESS mission and what is possible to do in crowded fields and with only limited sectors. Finally, our paper acts as an important proof-of-concept for the proposed ESA HAYDN mission \citep{Miglio2021}, which will specifically target dense environments, such as young open clusters, for the purpose of asteroseismology.

\section*{acknowledgements}

The authors gratefully acknowledge UK Research and Innovation (UKRI) in the form of a Frontier Research grant under the UK government’s ERC Horizon Europe funding guarantee (SYMPHONY; PI Bowman; grant number: EP/Y031059/1), and a Royal Society University Research Fellowship (PI Bowman; grant number: URF{\textbackslash}R1{\textbackslash}231631). J.B. is supported by an NWO Veni fellowship (VI.Veni.242.199). H.S. has received financial support from the Flemish Government under the long-term structural Methusalem funding program by means of the project SOUL: Stellar evolution in full glory, grant METH/24/012 at KU Leuven. Based on observations made with the Mercator Telescope (\url{https://www.mercator.iac.es}), operated on the island of La Palma by the Flemish Community, at the Spanish Observatorio del Roque de los Muchachos of the Instituto de Astrofísica de Canarias, and obtained with the HERMES spectrograph, which is supported by the Research Foundation - Flanders (FWO), Belgium, the Research Council of KU Leuven, Belgium, the Fonds National de la Recherche Scientifique (F.R.S.-FNRS), Belgium, the Royal Observatory of Belgium, the Observatoire de Genève, Switzerland and the Thüringer Landessternwarte Tautenburg, Germany. This research has used data obtained at the Mercator Observatory which receives funding from the Research Foundation – Flanders (FWO) (grant agreements I000325N and I000521N). This work presents results from the European Space Agency space mission Gaia. Gaia data are being processed by the Gaia Data Processing and Analysis Consortium (DPAC). Funding for the DPAC is provided by national institutions, in particular the institutions participating in the Gaia multilateral agreement.

\section*{Data Availability}

The TESS data used in this work are publicly available from the MAST archive: \url{https://archive.stsci.edu/missions-and-data/tess}.
The {\sc Hermes} spectra used in this work are available from the archive: \url{https://www.mercator.iac.es/instruments/hermes/archive/}.
The GAIA data used in this work are publicly available via the Gaia website: \url{https://gea.esac.esa.int/archive/}.
The Padova database of stellar evolutionary tracks and isochrones are publicly available via the website: \url{https://stev.oapd.inaf.it/cgi-bin/cmd}.
The software \textsc{lightkurve} and \textsc{period04}, used in this work for light-curve extraction and frequency analysis, respectively, are publicly available at \url{https://lightkurve.github.io/lightkurve/} and \url{https://www.period04.net/}.
This research has made use of the following software packages: {\tt matplotlib} \citep{Hunter_2007}, {\tt astropy} \citep{astropy1, astropy2, astropy3}, {\tt scipy} \citep{2020SciPy-NMeth}, {\tt pyAstronomy} \citep{pya}, {\tt TLUSTY} \citep{tlusty_1995, tlusty_2007}.
This research has made use of the SIMBAD database, operated at CDS, Strasbourg, France; the SAO/NASA Astrophysics Data System; and the VizieR catalogue access tool, CDS, Strasbourg, France. 

For the purpose of open access, the authors have applied a CC BY licence to the author-accepted manuscript version: \url{https://arxiv.org/abs/TBD}. Data products that support the results in this paper are publicly available via the Zenodo repository: \url{https://doi.org/10.5281/zenodo.20932892}.



\bibliographystyle{mnras}
\bibliography{bibliography} 




\clearpage
\newpage
\appendix

\section{Additional Tables}

The properties of the 82 B-type stars studied in this work are given in Table~\ref{tab:stars_list}. The extracted significant frequecies are given in Table~\ref{tab:significant_frequencies}.

\begin{table*}
\scriptsize
\caption{Properties of the 82 B-type stars studied in this work, which include spectral type and binary status from \citetalias{Nardini}, TESS contamination ratio (values higher than 1 mean that the majority of the flux is coming from a contaminating source), and type of pulsator based on the TESS light curve.}
\begin{tabular}{llccccc}
\hline \hline
SIMBAD ID       & TIC ID    & V [mag] & Spectral Type       & Binary status & $C_{\rm ratio}$ & Class                   \\ \hline
\multicolumn{2}{l}{Cluster: $h$ and $\chi$ Persei} & \multicolumn{2}{l}{TESS Sectors: 18, 58, 85}     & \multicolumn{1}{l}{} & \multicolumn{1}{l}{} & \multicolumn{1}{l}{} \\ \cline{1-4}
BD+56 515       & 264730455 & 8.53    & B2 V      & Single        & 1.9             & $\beta$ Cep             \\
BD+56 566       & 348444454 & 10.16   & B1e V     & Single        & 0.12            & Be                      \\
NGC 884 2114 & 348444627 & 11.04   & B2 V      & Single        & 4.5             & $\beta$ Cep$^{\dagger}$            \\
NGC 884 2185 & 348607492 & 10.92   & B5 V      & SB1           & 0.51            & Rot Modulation          \\
NGC 884 2235 & 348607658 & 8.40    & B1 V      & Single        & 3.4             & $\beta$ Cep$^{\dagger}$             \\
NGC 884 2246 & 348607672 & 9.98    & B1V       & Single        & 7.0             & $\beta$ Cep$^{\dagger}$             \\
NGC 884 2255 & 348607731 & 10.71   & B2 V      & SB1           & 7.4             & $\beta$ Cep$^{\dagger}$             \\
NGC 884 2296 & 348607543 & 8.53    & B1 II-III & Single        & 0.71            & $\beta$ Cep             \\
NGC 884 2299 & 348607604 & 8.32    & B1 II-III & Single        & 1.4             & $\beta$ Cep$^{\dagger}$             \\
NGC 884 2311 & 348607572 & 9.40    & B0 V      & SB1           & 3.0             & EB                      \\
NGC 884 2392 & 348607862 & 10.71   & B2 V      & SB2           & 0.58            & constant                \\
HD 14052        & 264612788 & 7.79    & B1 II III & SB1           & 0.30            & $\beta$ Cep             \\
NGC 869 0622    & 264612967 & 10.89   & B2 II III & Single        & 1.0             & $\beta$ Cep             \\
NGC 869 0843    & 264730428 & 8.38    & B1 V      & SB1           & 0.59            & $\beta$ Cep             \\
NGC 869 0864    & 264730166 & 9.16    & B2 V      & Single        & 0.87            & $\beta$ Cep             \\
NGC 869 0929    & 264730591 & 10.32   & B2 V      & Single        & 2.2             & $\beta$ Cep$^{\dagger}$             \\
NGC 869 0936    & 264730544 & 10.44   & B0 V      & Single        & 5.5             & $\beta$ Cep$^{\dagger}$             \\
NGC 869 0978    & 264730588 & 10.68   & B2 II III & Single        & 4.5             & -                       \\
NGC 869 0980    & 264730498 & 9.75    & B0 V      & SB1           & 7.2             & -                       \\
NGC 869 0992    & 264730502 & 10.00   & B1 II III & Single        & 11.5            & -                       \\
NGC 869 1004    & 264730535 & 10.91   & B2 V      & SB1           & 12.9            & -                       \\
NGC 869 1078    & 264730358 & 9.82    & B1 V      & Single        & 16.9            & -                       \\
NGC 869 1085    & 264730539 & 10.47   & B1 V      & SB1           & 5.5             & -                       \\
NGC 869 1116    & 264730427 & 9.29    & B1 V      & Single        & 10.1            & -                       \\
NGC 869 1132    & 264730392 & 8.48    & B2 II III & Single        & 3.4             & -                       \\
NGC 869 1133    & 264730365 & 9.04    & B1 II-III & Single        & 5.1             & -                       \\ \hline
\multicolumn{2}{l}{Cluster: NGC 457}               & \multicolumn{2}{l}{TESS Sectors: 18, 58}         & \multicolumn{1}{l}{} & \multicolumn{1}{l}{} & \multicolumn{1}{l}{} \\ \cline{1-4}
NGC 457 6       & 332499085 & 10.60   & B3e V     & SB1           & 3.1             & Be$^{\dagger}$          \\
NGC 457 7       & 332499065 & 10.95   & B5 V      & SB1           & 2.5             & -                       \\
NGC 457 8       & 332499097 & 10.02   & B1 V      & SB1           & 0.97            & $\beta$ Cep             \\
NGC 457 14      & 332499244 & 10.20   & B1e V     & SB1           & 2.4             & Be$^{\dagger}$          \\
NGC 457 19      & 332499278 & 9.51    & B1 II-III & SB1           & 0.67            & $\beta$ Cep             \\
NGC 457 33      & 332498935 & 10.33   & B2 V      & Single        & 0.25            & $\beta$ Cep             \\
NGC 457 34      & 332499026 & 10.74   & B5 V      & Single        & 2.3             & hybrid                  \\
NGC 457 37      & 332499008 & 9.83    & B2 V      & SB1           & 0.86            & EB                      \\
NGC 457 54      & 332499331 & 10.18   & B2 V      & SB1           & 0.19            & $\beta$ Cep             \\
NGC 457 85      & 332499083 & 10.78   & B5 V      & SB2           & 0.19            & EB                      \\
NGC 457 91      & 332499409 & 11.30   & B5e V     & SB1           & 0.38            & Be                      \\
NGC 457 100     & 332499478 & 10.61   & B2 V      & SB1           & 11.8            & - \\ 
NGC 457 120     & 332498822 & 9.93    & B0 V      & Single        & 0.14            & constant                \\
NGC 457 128     & 332489861 & 9.72    & B2e V     & Single        & 0.06            & Be                      \\
NGC 457 153     & 332489770 & 9.48    & B1e V     & Single        & 0.02            & Be                      \\
NGC 457 154     & 332489761 & 11.18   & B5 V      & SB1           & 0.30            & hybrid$^{\dagger}$                  \\
NGC 457 198     & 332490249 & 9.46    & B2e V     & Single        & 0.13            & Be                      \\
NGC 457 275     & 332489539 & 9.85    & B3e V     & Single        & 0.02            & Be                      \\
HD 236695       & 332497979 & 9.50    & B2 II III & Single        & 0.04            & $\beta$ Cep             \\ \hline
\multicolumn{2}{l}{Cluster: NGC 581}               & \multicolumn{2}{l}{TESS Sectors: 18, 25, 58, 85} & \multicolumn{1}{l}{} & \multicolumn{1}{l}{} & \multicolumn{1}{l}{} \\ \cline{1-4}
NGC 581 49      & 389771251 & 11.76   & B2e V     & Single        & 1.2             & Be                      \\
NGC 581 35      & 389771025 & 10.45   & B3 V      & SB1           & 0.19            & $\beta$ Cep             \\
NGC 581 59      & 389771267 & 11.45   & B3 V      & SB1           & 1.9             & $\beta$ Cep$^{\dagger}$             \\
NGC 581 70      & 389771358 & 11.74   & B5 V      & Single        & 22.3            & hybrid$^{\dagger}$      \\
NGC 581 73      & 389771312 & 10.59   & B5 V      & Single        & 1.5             & hybrid$^{\dagger}$      \\
NGC 581 162     & 389770884 & 11.22   & B2 V      & Single        & 0.23            & $\beta$ Cep             \\
NGC 581 111     & 389974549 & 11.84   & B5 V      & Single        & 0.30            & -                       \\
NGC 581 122     & 389771239 & 11.00   & B2 V      & SB1           & 36.1            & $\beta$ Cep$^{\dagger}$ \\
TYC 4031 2100 1 & 389766903 & 11.65   & B2 V      & Single        & 0.19            & -                       \\
BD+59 273       & 389771132 & 9.03    & B1 V      & Single        & 1.0             & $\beta$ Cep$^{\dagger}$ \\
EM* GGA 52      & 389771144 & 11.22   & B3 V      & SB1           & 0.73            & EB                      \\
EM* GGA 54      & 389771271 & 11.27   & B0e V     & Single        & 5.0             & Be                      \\
EM* GGA 56      & 389771456 & 11.35   & B5e V     & Single        & 0.50            & Be                      \\
V* V1122 Cas    & 389771376 & 9.71    & B3e V     & SB1           & 9.3             & -                       \\ \hline
\end{tabular}
\label{tab:stars_list}
\end{table*}

\begin{table*}
\scriptsize
\contcaption{}
\begin{tabular}{llccccc}
\hline \hline
SIMBAD ID    & TIC ID    & V {[}mag{]} & Spectral Type    & Binary status & $\rm C_{ratio}$ & Class                   \\ \hline
\multicolumn{2}{l}{Cluster: NGC 1960}               & \multicolumn{2}{l}{TESS Sectors: 19, 43, 44, 45, 59, 71, 73}         & \multicolumn{1}{l}{} & \multicolumn{1}{l}{} & \multicolumn{1}{l}{} \\ \cline{1-4}
NGC 1960 003 & 115722242  &   10.72      & B7 V   &   Single      &        2.1       &  -         \\
NGC 1960 008 & 115722132 & 9.36        & B2 V   & SB2           & 1.3                 & Rot Modulation          \\
NGC 1960 009 & 115722123 & 9.13        & B1 V   & SB1           & 0.97                & -          \\
NGC 1960 013 & 115722060  &   10.80  & B7 V   & Single &   1.3            &     -         \\
NGC 1960 014 & 115722047 & 10.68   & B7 V  &   Single   &  1.3            &     -         \\
NGC 1960 016 & 115722162 & 8.79        & B2 V   & SB1           & 0.15                & Rot Modulation             \\
NGC 1960 018 & 115722189 & 10.78       & B5 III & Single        & 2.2                 & hybrid$^{\dagger}$             \\
NGC 1960 021 & 115722282 & 9.60        & B2 V   & Single        & 1.9                 & $\beta$ Cep$^{\dagger}$             \\
NGC 1960 023 & 115722303 & 8.96        & B3 V   & Single        & 0.45                & $\beta$ Cep \\
NGC 1960 027 & 115722156 & 9.58        & B2e V  & SB1           & 0.19                & Be                      \\
NGC 1960 038 & 115721955 & 9.92        & B3 V   & Single        & 0.20                & $\beta$ Cep             \\
NGC 1960 047 & 115559839 & 10.44       & B5e V  & Single        & 0.64                & Be$^{\dagger}$                      \\
NGC 1960 048 & 115559789 & 9.36        & B3 V   & Single        & 0.31                & $\beta$ Cep             \\
NGC 1960 061 & 115722319 & 9.09        & B2 V   & Single        & 0.07                & $\beta$ Cep \\ 
NGC 1960 081 & 115559920 & 9.99        & B2 V   & Single        & 0.16                & SPB  \\ 
NGC 1960 086 & 115559710 &  10.65  & B7 V  & Single        & 0.44                 &      hybrid        \\
NGC 1960 087 & 115559692 & 10.65       & B5 V   & Single        & 0.25                &  hybrid \\ 
NGC 1960 091 & 115559630 & 10.37       & B5 V   & Single        & 0.20                &  SPB  \\
NGC 1960 092 & 115559595 &   10.96 & B9 V & Single        &  0.42             &    hybrid $\delta$ Scuti           \\
NGC 1960 101 & 115722396 & 9.14        & B2e V  & Single        & 0.03                & Be                      \\
NGC 1960 109 & 115642252 & 10.70       & B3 V   & SB1           & 0.85                & -               \\
NGC 1960 134 & 115730751 & 10.39       & B2 V   & SB1           & 0.06                & heartbeat               \\
NGC 1960 138 & 115642245 & 8.94        & B2 V   & Single        & 0.19                & $\beta$ Cep             \\ \hline
\end{tabular}
\label{tab:star_list_continue}
     \begin{tablenotes}
    \item \textbf{Notes. } $^{\dagger}$Stars for which we tentatively classify the pulsation class, but the contamination level is high.
    \end{tablenotes}
\end{table*}

\begin{table}
\scriptsize
\centering
\caption{Names, significant frequencies, amplitudes and phases of the 22 $\beta$~Cep, SPBs and hybrids identified in the four different clusters.
}
\begin{tabular}{llccc}
\hline \hline
simbad ID    & TIC ID    & $f$ [d$^{-1}$]  & A [mmag]         & $\phi$      \\ 
\hline
BD+56 515    & 264730455 & 0.83409(8) & 0.82(2)  & 0.15 $\pm$ 0.23  \\
             &           & 1.38075(6) & 1.09(2)  & 0.62 $\pm$ 0.17  \\
             &           & 2.12101(6) & 1.13(2)  & 0.76 $\pm$ 0.17  \\
NGC 884 2296 & 348607543 & 0.74476(7) & 3.63(7)  & 0.86 $\pm$ 0.20  \\
HD 14052     & 264612788 & -          &    -      &    -              \\
NGC 869 0622   & 264612967 & 0.38338(4) & 1.20(1)  & 0.32 $\pm$ 0.13  \\
             &           & 5.8236(3)  & 0.16(1)  & 0.14  $\pm$ 0.97 \\
NGC 869 0843   & 264730428 & 2.1400(2)  & 0.21(1)  & 0.86 $\pm$ 0.54  \\
             &           & 3.1849(3)  & 0.14(1)  & 0.80 $\pm$ 0.80  \\
             &           & 4.50047(6) & 0.70(1)  & 0.39  $\pm$ 0.16 \\
             &           & 9.1816(3)  & 0.16(1)  & 0.90  $\pm$ 0.75 \\
NGC 869 0864   & 264730166 & 0.22612(5) & 1.58(2)  & 0.07 $\pm$ 0.11  \\
             &           & 1.25688(8) & 0.75(2)  & 0.05 $\pm$ 0.23  \\
             &           & 7.0492(5)  & 0.13(2)  & 0.47 $\pm$ 1.30  \\
             &           & 14.1087(4) & 0.14(2)  & 0.03 $\pm$ 1.22  \\
NGC 457 8    & 332499097 & 0.44901(3) & 2.41(2)  & 0.65 $\pm$ 0.10  \\
             &           & 5.34021(1) & 8.64(2)  & 0.69 $\pm$ 0.03  \\
             &           & 5.4331(3)  & 0.28(2)  & 0.21 $\pm$ 0.85  \\
NGC 457 19   & 332499278 & 0.80020(7) & 1.46(3)  & 0.05 $\pm$ 0.22  \\
NGC 457 33   & 332498935 & 0.18151(5) & 2.77(3)  & 0.99 $\pm$ 0.14  \\
NGC 457 34   & 332499026 & 2.8012(1)  & 0.64(2)  & 0.96 $\pm$ 0.36  \\
             &           & 5.34038(3) & 2.59(2)  & 0.17 $\pm$ 0.09  \\
NGC 457 54   & 332499331 & 1.0307(1)  & 0.34(1)  & 0.24 $\pm$ 0.42  \\
HD 236695    & 332497979 & 0.38916(3) & 3.74(3)  & 0.98 $\pm$ 0.09  \\
             &           & 0.47512(5) & 2.27(3)  & 0.70 $\pm$ 0.15  \\
NGC 581 35   & 389771025 & 0.66973(7) & 0.64(2)  & 0.96 $\pm$ 0.27  \\
             &           & 1.43361(6) & 0.86(2)  & 0.93 $\pm$ 0.20  \\
NGC 581 162  & 389770884 & 1.0915(1)  & 0.52(2)  & 0.79 $\pm$ 0.45  \\
             &           & 1.19255(6) & 1.00(2)     & 0.74 $\pm$ 0.23  \\
             &           & 2.7743(2)  & 0.28(2)  & 0.41 $\pm$ 0.81  \\
NGC 1960 23  & 115722303 & 0.61977(5) & 0.62(9)  & 0.88 $\pm$ 0.13  \\
             &           & 1.7085(2)  & 0.17(9)  & 0.06 $\pm$ 0.47  \\
             &           & 1.8380(2)  & 0.19(9)  & 0.90  $\pm$ 0.44 \\
NGC 1960 38  & 115721955 & 0.6994(1)  & 0.22(7)  & 0.6 $\pm$ 0.32   \\
             &           & 0.82496(9) & 0.31(7)  & 0.49 $\pm$ 0.23  \\
             &           & 1.5531(2)  & 0.19(7)  & 0.30 $\pm$ 0.38  \\
             &           & 2.0437(2)  & 0.18(8)  & 0.96 $\pm$ 0.40  \\
             &           & 2.1954(2)  & 0.17(8)  & 0.67 $\pm$ 0.42  \\
             &           & 2.30293(4) & 0.78(8)  & 0.16 $\pm$ 0.09  \\
             &           & 19.2089(6) & 0.05(7)  & 0.81 $\pm$ 1.42  \\
             &           & 20.4141(4) & 0.07(7)  & 0.43 $\pm$ 0.98  \\
             &           & 27.4541(3) & 0.09(7)  & 0.83 $\pm$ 0.77  \\
             &           & 28.0289(6) & 0.05(7)  & 0.18 $\pm$ 1.51  \\
NGC 1960 48  & 115559789 & 0.4437(1)  & 0.16(6)  & 0.94 $\pm$ 0.35  \\
             &           & 0.5567(2)  & 0.12(6)  & 0.01 $\pm$ 0.47  \\
             &           & 2.3191(3)  & 0.09(6)  & 0.87 $\pm$ 0.64  \\
             &           & 2.90732(3) & 0.07(6) & 0.58 $\pm$ 0.79  \\
             &           & 3.5817(1)  & 0.18(6)  & 0.33 $\pm$ 0.32 \\
             \hline
\end{tabular}\label{tab:significant_frequencies}
\end{table}

\begin{table}\label{tab:significant_frequencies_continue}
\scriptsize
\centering
\contcaption{}
\begin{tabular}{llccc}
\hline \hline
simbad ID    & TIC ID    & $f$ [d$^{-1}$]  & A [mmag]         & $\phi$      \\ 
\hline
NGC 1960 61 & 115722319 & 0.41092(7) & 0.75(1) & 0.65 $\pm$ 0.19  \\
            &           & 1.24597(7) & 0.82(2) & 0.54 $\pm$ 0.17  \\
            &           & 1.2997(2)  & 0.37(2) & 0.09 $\pm$ 0.42  \\
            &           & 1.34674(1) & 0.48(2) & 0.77 $\pm$ 0.30  \\
            &           & 1.44330(1) & 0.44(2) & 0.67 $\pm$ 0.33  \\
            &           & 2.4956(3)  & 0.21(1) & 0.58 $\pm$ 0.70  \\
            &           & 2.6379(2)  & 0.25(1) & 0.82 $\pm$ 0.57  \\
            &           & 2.8708(4)  & 0.16(1) & 0.25 $\pm$ 0.88  \\
            &           & 5.0301(5)  & 0.11(1) & 0.02 $\pm$ 1.26  \\
            &           & 5.0976(5)  & 0.12(1) & 0.31 $\pm$ 1.14   \\
            &           & 6.4457(3)  & 0.19(1) & 0.79 $\pm$ 0.76  \\
            &           & 6.6540(6)  & 0.09(1) & 0.21 $\pm$ 1.57  \\
            &           & 7.2726(8)  & 0.07(1) & 0.66 $\pm$ 1.99  \\
            &           & 7.8251(4)  & 0.14(1) & 0.98 $\pm$ 1.02  \\
            &           & 10.2245(5) & 0.11(1) & 0.99 $\pm$ 1.24   \\
            &           & 10.3646(4) & 0.16(1) & 0.38 $\pm$ 0.90   \\
            &           & 10.5385(6) & 0.09(1) & 0.59 $\pm$ 1.51 \\
NGC 1960 87 & 115559692 & 1.4779(1)  & 0.35(2)  & 0.83 $\pm$ 0.26 \\
            &           & 2.8596(2)  & 0.207(8) & 0.57 $\pm$ 0.38 \\
            &           & 2.9619(2)  & 0.185(9) & 0.78 $\pm$ 0.43 \\
            &           & 3.0333(3)  & 0.122(8) & 0.28 $\pm$ 0.65 \\
            &           & 3.2161(3)  & 0.103(8) & 0.16 $\pm$ 0.77 \\
            &           & 4.4331(4)  & 0.10(1)  & 0.77 $\pm$ 0.89 \\
            &           & 5.82025(8) & 0.387(8) & 0.21 $\pm$ 0.20 \\
            &           & 5.9174(5)  & 0.060(8) & 0.10 $\pm$ 1.31 \\
            &           & 6.03537(4) & 0.864(8) & 0.20 $\pm$ 0.09 \\
            &           & 6.0907(6)  & 0.053(8) & 0.94 $\pm$ 1.50 \\
            &           & 6.3131(2)  & 0.175(8) & 0.20 $\pm$ 0.45 \\
NGC 1960 91 & 115559630 & 0.08866(6)  & 0.61(1) & 0.08 $\pm$ 0.16 \\
            &           & 2.4550(1)   & 0.39(1) & 0.27 $\pm$ 0.25 \\
            &           & 2.4935(2)   & 0.16(1) & 0.77 $\pm$ 0.61 \\
            &           & 2.54095(8)  & 0.55(1) & 0.36 $\pm$ 0.20 \\
            &           & 2.5952(2)   & 0.18(1) & 0.76 $\pm$ 0.56 \\
            &           & 2.630129(1) & 2.86(1) & 0.49 $\pm$ 0.03 \\
            &           & 2.718156(1) & 4.38(1) & 0.71 $\pm$ 0.02 \\
            &           & 2.78065(8)  & 0.50(1) & 0.80 $\pm$ 0.20 \\
            &           & 2.8577(3)   & 0.12(1) & 0.68 $\pm$ 0.83 \\
            &           & 3.1156(2)   & 0.25(1) & 0.11 $\pm$ 0.38 \\
            &           & 5.1705(3)   & 0.14(1) & 0.29 $\pm$ 0.70 \\
            &           & 5.2317(3)   & 0.14(1) & 0.36 $\pm$ 0.73 \\
            &           & 5.3221(1)   & 0.32(1) & 0.43 $\pm$ 0.32 \\
            &           & 5.3566(3)   & 0.14(1) & 0.70 $\pm$ 0.69 \\
            &           & 5.43484(4)  & 0.89(1) & 0.59 $\pm$ 0.11 \\
            &           & 5.5002(2)   & 0.20(1) & 0.64 $\pm$ 0.51 \\
            &           & 5.5642(4)   & 0.11(1) & 0.67 $\pm$ 0.89 \\
            &           & 5.6842(3)   & 0.15(1) & 0.09 $\pm$ 0.64 \\
            &           & 5.7456(3)   & 0.14(1) & 0.83 $\pm$ 0.70 \\
            &           & 5.8340(4)   & 0.10(1) & 0.19 $\pm$ 0.94 \\
            &           & 8.0266(4)   & 0.10(1) & 0.27 $\pm$ 0.95 \\
            &           & 10.69714(1) & 0.35(1) & 0.21 $\pm$ 0.27 \\
            &           & 10.9122(2)  & 0.21(1) & 0.27 $\pm$ 0.47 \\
            &           & 16.0053(8)  & 0.05(1) & 0.33 $\pm$ 1.98 \\
            &           & 16.1415(7)  & 0.06(1) & 0.09 $\pm$ 1.69 \\
NGC 1960 138 & 115642245 & 0.93352(4) & 0.72(8) & 0.84 $\pm$ 0.11 \\
             &           & 1.2550(2)  & 0.15(8) & 0.96 $\pm$ 0.53 \\
             &           & 1.8663(2)  & 0.19(8) & 0.59 $\pm$ 0.42 \\
             &           & 2.8764(1)  & 0.27(8) & 0.48 $\pm$ 0.29 \\
             &           & 2.9804(4)  & 0.09(8) & 0.48 $\pm$ 0.88 \\
             &           & 3.1401(2)  & 0.13(8) & 0.05 $\pm$ 0.61 \\
             &           & 4.6486(4)  & 0.77(8) & 0.98 $\pm$ 1.02 \\
             &           & 4.9161(2)  & 0.13(8) & 0.28 $\pm$ 0.60 \\
             \hline
\end{tabular}
\end{table}

\section{Notes on individual systems}\label{appendix:a}
\subsection{Pulsating Be stars}\label{appendix:individual_Be}
\textit{NGC~1960~101:} the light curve and frequency spectrum of this Be star are an example of perhaps the most extreme outburst within the sample.
The light curve is shown in the first row of Fig.~\ref{fig:green_be}, and displays strong sector-to-sector variability. In the three consecutive TESS sectors shown, each exhibits a prominent increase in brightness near the beginning. We investigated potential instrumental causes and possible contamination from bright nearby sources, but found no evidence for either. Given the classification of this target as a Be star based on strong \ion{H$\alpha$}{} emission, the large photometric variability is likely related to the presence of a decretion disc and repeated outbursts \citep[][]{Huat_2009, Labadie_Bartz_2022}. The frequency spectrum is dominated by an upward trend at very low frequency (below 1~d$^{-1}$) caused by the intermittent quasi-periodic outbursts. For this reason, there are no dominant gravity-mode frequencies or frequency groups.
We note that, other than PCA detrending, no additional spline detrending was performed for the TESS light curves of this star due to its large-scale variability. Regarding the spectra, the bottleneck shape of the \ion{H$\alpha$}{} emission line, together with its low projected rotational velocity, means that this Be star is likely viewed close to pole-on \citep[][]{Sigut_2023}. The amplitude of the emission line is extremely high, and there is a decrease in flux between the first spectra taken in 2016 and the last one taken in 2024, indicating that the decretion disc is transient. 

\textit{BD+56~566:} This is the only pulsating Be star in $h$ and $\chi$~Per and exhibits the strongest long-term photometric variability across all three available TESS sectors during 2019 (sector 18), 2022 (sector 58) and 2024 (sector 85). The frequency spectrum of the non-detrended light curve is dominated by the long-term trends, but multi-periodic gravity-modes and frequency groups at about 1~d$^{-1}$ and 2.5~d$^{-1}$ are evident in the frequency spectrum of the detrended light curve (see second row of Fig.~\ref{fig:green_be}).
The two available {\sc Hermes} spectra were obtained in March 2024 with a separation of only 2~d, which is too short to reveal any outburst signatures via changes in the emission lines. The unequal peak heights of \ion{H$\alpha$}{} might be caused by the V/R cycle, but the lack of spectral coverage does not allow us to confirm if there is spectroscopic variability. 

\textit{NGC~457~198:} This Be star has the longest-period pulsations and smaller amplitude changes due to beating compared to the rest of the sample. There are four dominant frequency groups in the frequency spectrum (see third row of Fig.~\ref{fig:green_be}), and all of them are below 2~d$^{-1}$. The first {\sc Hermes} spectrum was taken in 2016 and shows lower amplitude \ion{H$\alpha$}{} emission compared to the additional 17 spectra, which show little-to-no variability until the last spectrum taken in 2022. 

\textit{EM$^*$~GGA~54:} The {\sc Hermes} spectra show signatures of a V/R cycle, but with only four epochs, we lack the coverage to establish the variability cycle. The largest difference in \ion{H$\alpha$}{} emission is observed between the last two spectra, which were obtained one year apart.

\textit{NGC~457~275:} Two of the {\sc Hermes} spectra for this star were obtained contemporaneously with TESS sector 18 in 2019. No long-term trends are seen in the light curve, and no variability is detected in the {\sc Hermes} spectra across this 28-d time span (see fourth row of Fig.~\ref{fig:green_be}, which suggests that the star did not undergo any significant outbursts in this time. The additional {\sc Hermes} spectra span three years and do not show large variability in \ion{H$\alpha$}{}. Moreover, the shape of \ion{H$\alpha$}{} indicates that the star is seen almost pole-on.

\textit{NGC~457~153:} The bottle-neck shape of \ion{H$\alpha$}{} in this Be star indicates an inclination angle of around 40~deg \citep[][]{Sigut_2023}, and there are no signs of variability across the 4-yr time span. As shown in the sixth row of Fig.~\ref{fig:green_be}, there are no long-term trends in the two TESS sectors available (sectors 18 and 58). The frequency spectra show two peaks at about 0.8 and 1.5~d~$^{-1}$, which are likely frequency groups.

\textit{NGC~457~128:} There is no evidence of long-term photometric variability in the two available TESS sectors (18 and 58). However, the multi-epoch {\sc Hermes} spectra spanning almost 3~yr show time-dependent emission in the Balmer lines (see seventh row of Fig.~\ref{fig:green_be}). Several {\sc Hermes} spectra obtained between 2017 and 2020 reveal emission features only in the four epochs observed in 2019, suggesting a possible mass-ejection event that supplied a decretion disc for roughly five months (from August to November 2019). By January 2020, the emission had disappeared and \ion{H$\alpha$}{} returned to being in absorption. Sector 18 of the TESS mission occurred in November 2019, but no brightening or fading occurred during this time,  and only multi-periodic gravity-mode pulsations are evident. The small amplitude of the \ion{H$\alpha$}{} emission (around 0.2 in normalised flux) suggests that the mass-ejection-induced brightness increase may not be sufficient to produce a visible flux increase in the broad-band TESS light curve.

\textit{NGC~457~91:} This Be star is another example of long-term photometric variability in both available TESS sectors.
In the non-detrended sector 58 TESS light curve, which is shown in the eighth row of Fig.~\ref{fig:green_be}, there are small fading and brightening events. While the frequency groups remain consistent between the two TESS sectors, the amplitudes of the pulsation modes are lower in sector 18 compared to sector 58. One of the {\sc Hermes} spectra was taken in the same month of sector 58, but no variability in \ion{H$\alpha$}{} is detected compared to the other available spectra. The {\sc Hermes} spectra span from 2017 to 2020 and show no strong variability apart from a slight decrease in amplitude of H$\alpha$ between the first and second year of observations.

\textit{NGC~581~49:} This Be star has the smallest pulsation amplitudes in the light curve and frequency spectrum in our sample. Sector 58 shows long-term photometric variability (see ninth row of Fig.~\ref{fig:green_be}), but it depends on which aperture mask is used, so this is deemed to be instrumental and not astrophysical in origin.

\textit{NGC~960~27:} There are seven TESS sectors available for this star, with the three consecutive sectors 43, 44, and 45 shown in the tenth row of Fig.~\ref{fig:green_be}. The larger amount of TESS data is reflected in the dense frequency groups in the frequency spectrum, with the dominant gravity-mode pulsations of about 2.2~d$^{-1}$. 
The twelve {\sc Hermes} spectra obtained between 2016 and 2024 show a steady increase in the amplitude of \ion{H$\alpha$}{}, indicating continuous disc growth during this time.

\textit{EM$^*$~GGA~56:} This star has the highest projected rotational velocity of all the Be stars in our sample with $v\,\sin\,i = 360$~km~s$^{-1}$. It is also among those with the largest number of frequency groups (five), with the highest-frequency one detected at 7~d$^{-1}$, as shown in the eleventh row of Fig.~\ref{fig:green_be}. The eight available {\sc Hermes} spectra show no variability in \ion{H$\alpha$}{} across 516~d, although the last spectrum was obtained in 2021, which pre-dates the TESS data.

\subsection{Photometric binaries}
\label{appendix:individual_binaries}
\textit{NGC~884~2311:} (also known as V621~Per) The TESS data are shown in the first row of Fig.~\ref{fig:green_eb}. It was previously identified as an EB by \citet{Laur}, but the system remains to be modelled to extract dynamical masses and radii. 
NGC~884~2311 is the longest-period EB in our sample with an orbital period of $25.5237\pm0.0054$~d and an eccentricity of $0.26821\pm0.00006$ determined from spectroscopic RV analysis in \citetalias{Nardini}. The primary eclipses occur near the beginning and end of the 27-d period for the TESS sector 58, and the asymmetric time differences between secondary and primary eclipses support the high eccentricity. 
Additionally, one or both of the components is a pulsating star, with pulsating EBs invaluable for constraining stellar structure and evolution theory (see \citealt{Southworth_Bowman_2026}).
NGC~884~2311 suffers from a moderate level of contamination, so whilst the spectroscopic and photometric solutions for binarity are in agreement, it is not possible with current data sets to ascertain the origin of the pulsations.

\textit{NGC~457~37:} (also known as BD+57~249) \citet{Zhang} were the first to identify the system as an EB and determine the orbital period of $4.44768\pm0.00008$~d using ground-based two-colour photometry. 
This is consistent with the spectroscopic binary period determined in \citetalias{Nardini}, which is $P_{\rm spec}=4.4475\pm0.0003$~d, as well as the photometric period found in this work of $P_{\rm phot}=4.447500\pm0.000001$~d. The similar eclipse depths of NGC~457~37 indicate it contains two stars of similar brightness in a circular orbit. 

\textit{NGC~457~85:} (also known as V765~Cas) This is the only EB in our sample that is also SB2. It was first identified as an EB by \citet{alfonso} 
and it was later modelled by \citet{Breeden_2024} using photometry from different instruments (including TESS), as well as undergoing isochrone fitting. Based on the shape of the light curve and the location of the star on the CMD, the authors classified it as a $\beta$~Lyr-type binary. 
The analysis of \citet{Breeden_2024} confirms the period of 1.715779~d from \citet{alfonso}, which is also consistent with the spectroscopic binary period of $P_{\rm spec}=1.7160 \, \pm \, 0.0002$~d from \citetalias{Nardini} and $P_{\rm phot}=1.714099 \, \pm \, 0.000004$~d determined in this work.

\textit{EM$^{*}$ GGA 52:} This is also previously identified as an EB by \citet{Laur} but no modelling was performed. The spectroscopic period of $P_{\rm spec}=2.2507 \pm 0.0002$~d from \citetalias{Nardini} {\bf is} half the photometric period of $P_{\rm phot}=4.502164 \pm 0.000003$~d derived in this work. The light curve in Fig.~\ref{fig:green_eb} demonstrates that the spectroscopic solution is an alias of the true photometric binary period. 

\textit{NGC 1960 134:} This is the only photometric binary in our sample that was not previously known in the literature. 
It is a short-period binary system with $P_{\rm spec}=8.8600 \pm 0.0021$~d and high eccentricity of $e= 0.54431 \pm 0.00015$ according to \citetalias{Nardini}. The TESS light curve is shown in the bottom row of Fig.~\ref{fig:green_eb}, and shows periodic brightening on the same time scale as the spectroscopic orbital period. The high eccentricity is likely sufficient to cause significant tidal distortion at periastron passage and give rise to a periodic brightening, hence it is an eccentric ellipsoidal variable. 
Binary tidal forces can also induce pulsations at integer harmonics of the orbital period \citep[see ][]{Southworth_Bowman_2026}. Once these harmonics are removed, it is possible to remove the binary signal from the light curve and inspect to see if there are signs of pulsation variability. Indeed, once removed, the light curve shows signs of pulsations, and the peaks in the frequency spectrum are between 2 and 3~d$^{-1}$ and are likely caused by pulsations. This suggests that one of the components of the binary system is either an SPB or a $\beta$ Cep star.

\subsection{Rotational modulation stars}\label{appendix:individual_rot_mod}
\textit{NGC~1960~8:} This is one of three SB2 systems identified in \citetalias{Nardini} and as the spectra were not disentangled no $v\,\sin\,i$ measurement could be obtained. The spectroscopic and photometric periods of $P_{\rm spec} = 2.4973 \pm 0.0002$~d and $P_{\rm phot} = 2.49853 \pm 0.00001$~d are identical within their respective confidence intervals, which demonstrates that the star with rotational modulation is synchronised with the orbit.

\textit{NGC~1960~16:} This is an SB1 system that lies on the bisector in Fig.~\ref{fig:1srharm_orbper} such that its spectroscopic and dominant photometric periods are consistent, but it is not classified as a photometric binary. The sinusoidal shape of the light curve and of the phase-folded light curve are typical of a star with rotational modulation, but since it is not perfectly periodic, this could mean either variable chemical abundance patterns or multi-periodic pulsations. The absorption lines in the spectra did not show line profile variability apart from RV variability due to binarity, which exceeds 43~km~s$^{-1}$. If this is a pulsator, then it seems rather coincidental that the dominant pulsation mode frequency is near the binary frequency. Whereas, if this is a star with rotational modulation, it is thus synchronised with the orbit.

\textit{NGC~884~2185:} This is the other SB1 system in our sample that lies on the bisector in Fig.~\ref{fig:1srharm_orbper}, but it is not classified as a photometric binary based on its TESS light curve in this work. The photometric and spectroscopic periods of $P_{\rm phot} = 1.45924 \pm 0.00004$~d and $P_{\rm spec} = 1.5483 \pm 0.0002$~d are similar but not compatible considering their respective confidence intervals.
The sinusoidal shape of the TESS light curve and especially of the phase-folded light curve, which are shown in Fig.~\ref{fig:rot_modulation}, is tentative for classifying this star as having rotational modulation. Since there is only one consecutive TESS sector available, this limits our frequency resolution. Moreover, the S/N ratios of the available {\sc Hermes} spectra are low compared to most of the stars in our sample, and the peak-to-peak RV variability is only $\sim29$~km~s$^{-1}$. Therefore, the argument that the star is rotationally synchronised with the orbital system and that the spectroscopic period is caused by chemical abundance anomalies is tentative.


\bsp	
\label{lastpage}
\end{document}